\documentstyle[preprint,eqsecnum,aps]{revtex}
\newcommand{\lymax}{\lambda_{\text{max}}}

\begin{document}
%%%%%%%%%%%%%%%%
%%%  Introduction %%%%%%
%%%%%%%%%%%%%%%%
\draft
%\preprint{STH-99-03}
\title{Periodicity Manifestations in the Turbulent Regime
of Globally Coupled Map Lattice
}
\author{Tokuzo Shimada and Kengo Kikuchi}
\address{Department of Physics, Meiji University\\
Higashi-Mita 1-1, Kawasaki, Kanagawa 214, Japan
}
\date{\today}
\maketitle
\begin{abstract}
We revisit the globally coupled map lattice (GCML).
We show that in the so called turbulent regime various periodic cluster attractor states 
are formed even though the coupling between the maps are very small relative 
to the nonlinearity in the element maps. 
Most outstanding is a maximally symmetric three cluster attractor 
in period three motion (MSCA) due to the foliation of the period three window 
of the element logistic maps. An analytic approach is proposed 
which explains successfully the systematics of various periodicity manifestations 
in the turbulent regime. The linear stability of the period three cluster attractors
is investigated. \\

\end{abstract}
%%%%%
\section{INTRODUCTION}

Recently there has been much progress in the study of synchronization of 
nonlinear maps \cite{ka,kb1,pikovskyb,kb2,kc,kd,sinha} and 
flows \cite{pecora,kowalski,noise,fns,pikovskya}.
This may lead to the clarification of the intelligence activity supposed to come from 
the synchronization among the neurons in the neural network. 
Especially the globally coupled map lattice (GCML) may be considered
as one of the basic models for the network systems expressing their 
characteristic limits. 
In its simplest form, all elements interact among themselves via their mean
field all to all with a common coupling, and each of the element is a simple logistic
map with a given nonlinearity.  It may be regarded as a natural extension of the 
spin-glass theories \cite{biology,spinglass} to the nonlinear dynamics.
Even though the simplest GCML has only two model parameters---the common nonlinearity 
parameter $a$ and the overall coupling $\varepsilon$---
it exhibits a rich variety of interesting phases on the parameter space 
corresponding to various forms of synchronization among the maps 
determined by the balance between the randomness specified by $a$ and 
the coherence by $\varepsilon$.

In this article we revisit the turbulent regime of GCML,
which is a regime in the parameter space with high $a$ and very small $\varepsilon$. 
The main interest in this regime has been so far focused to the so called hidden 
coherence \cite{kb1,pikovskyb}. 
It is a phenomenon that the fluctuation of the mean field of the
maps in evolution does not cease at large system size. 
The mean field distribution obeys the central limit theorem (CLT) 
but not the law of large number (LLN) \cite{kb1,kb2}.
We show that the dynamics of GCML in this regime is 
a foliation of that of the element logistic maps and that various periodic 
cluster attractors are formed even though the coupling between the maps 
is set very small.  We show that the regions which may be described 
by the hidden coherence do exist but that it is a very limited part of the parameter space. 

We organize our discussion in three parts. 
First, we present the results of an extensive phenomenological survey of this 
regime and list evidences of periodicity manifestations due to the periodic windows 
of the element logistic map. 
Most outstanding is the onset of period three cluster attractors.
The turbulent regime is, if we may say, 
a bizarre region with many faces---drastic periodicity manifestations as well 
as almost perfect randomness under the hidden coherence. 
At the periodic or quasi-periodic attractors, the mean field evolves controlled by
the scale of the cluster orbits and the LLN is naturally violated. 
We also present a remarkable data which shows that the GCML 
at the large system size acquires a high sensitivity to the periodic 
windows of the element map.
Second, we present an analytical approach which successfully explains the 
systematics of the periodicity manifestations.  We present a tuning condition
which limits the system parameters with which GCML cluster attractor states of 
a given periodicity may be formed.  
The key to obtain the condition is the introduction of the maximally symmetric 
cluster attractor (MSCA), which is a solution of minimum fluctuation in the mean field.
It corresponds to the known state of two clusters in opposite phase oscillation 
which is formed in the ordered phase of GCML \cite{ka}.  
We verify the validity of the condition in detail and show that the foliation is 
the governing dynamics of this regime.

Third, we show that the period three cluster attractors formed in the turbulent regime 
are linear stable and investigate how their stability changes by the coupling $\varepsilon$ 
and the population ratios. We in particular derive algebraically the $\varepsilon$ value 
for the formation of the most linearly stable bifurcated-MSCA (MSCA${}^*$).

The organization of this article is as follows. In Sec. II we briefly review various 
GCML phases and locate the turbulent regime in the parameter space. We then 
summarize the known facts of this regime. No originality is claimed here.
We then briefly compare them with our results.   
In Section III we present our phenomenological findings
including the period three MSCA and associated fewer cluster attractor.
In Sec. IV we present an analytic approach which explains successfully how the periodic windows 
of the element map control the GCML dynamics in the turbulent regime. 
In Sec. V we investigate the stability of the cluster attractors. We conclude in Sec. VI.

%%%%%%%%%%%%%%%%
%%% Section 2 %%%%%%%
%%%%%%%%%%%%%%%%
\section{GCML PHASE STRUCTURE AND THE TURBULENT REGIME: A REVIEW}

We study in this article the simplest GCML which is a system of $N$ maps evolving by
\begin{eqnarray}
 x_i(n+1) = (1-\varepsilon)f(x_i(n))+\varepsilon h(n), ~~~~( i =1, \cdots ,N), \label{global1}
\end{eqnarray}
and the mean field $h(n)$ of maps is defined as 
\begin{equation}
	h(n) \equiv \frac{1}{N}\sum_{i=1}^N f(x_i(n))=\frac{1}{N}\sum_{i=1}^N x_i(n+1). 
\label{meanfield}
\end{equation}
In the first step, all $x_i$ are simultaneously mapped by a nonlinear function $f$.  
The function $f$ could be distinct maps (heterogeneous GCML) but in this article we consider  
the simplest case that $f$ is a 
logistic map $f(x)=1-a x^2$ common to all variables (homogeneous GCML).
The nonlinearity of $f$ generally magnifies the variance among the maps.
The larger the parameter $a$ is, the more strongly the variance is enhanced. 
In the second, the maps undergo interaction  between themselves with a global 
coupling constant $\varepsilon$.  Here every $f(x_i)$ is pulled to their meanfield $h(n)$ 
at a fixed rate $1-\varepsilon$.
The larger the coupling $\varepsilon$ is,  the more strongly the maps are driven 
into synchronization. 

The model is endowed with various interesting phases under a subtle balance between the
two conflicting tendencies.   
The phase diagram in the $a, \varepsilon$ plane was explored by Kaneko  \cite{ka}.
Let us explain the phases choosing $a = 1.80$ for definiteness. 
This is far above the criticality $a=1.401\cdots$ to the chaos for a single logistic map.  
(i) For a sufficiently large $\varepsilon$ ($\ge 0.38)$ the maps are strongly bunched 
together in a cluster in the final attractor and evolve chaotically as a single logistic map. 
This is the {\it coherent phase}. 
(ii) For $\varepsilon=0.22-0. 30$ the interaction via the mean field can 
no longer exerts strong bunching and the final maps divide into two clusters. 
The maps in each cluster are still tightly synchronizing each other, and 
the two clusters mutually oscillate opposite in phase.   
This phase turns out as a solution of a minimum fluctuation in the mean field
and called as the {\it two-clustered ordered phase}.  
(iii) For smaller $\varepsilon$, the number of final clusters increases but it remains 
independent from the total number of maps in the system. 
The typical number of clusters at various $\varepsilon$ ranges is indicated 
in the phase diagram \cite{ka}.  
(iv) Finally,  for very small $\varepsilon$ the number of clusters is in general 
proportional to $N$. This region is called as the {\it turbulent regime}. 
This is the target region of our analysis.

It is known that in the turbulent regime maps evolve almost randomly at small lattice size 
$N$ and that there occurs a subtle correlation --- a hidden coherence --- at large $N$.
But as we show below there actually emerge drastic global periodic motions of maps 
if the $\varepsilon$ takes certain values for a given $a$. 
Let us first briefly review previous observations in the literature.

(i) The final state of GCML in this regime iterated from a random configuration 
consists of maps and tiny clusters, each moving chaotically due to high nonlinearity.
The number of elements (maps and clusters) is proportional to the number of whole maps 
in sharp contrast to the ordered regime \cite{kb1}. 

(ii) There emerges certain coherence between elements when the size $N$ is large \cite{kb1,kb2}. 
If $x_i(n)~(i=1,\cdots N)$ are really independent random variables following a common 
probability distribution, the mean squared deviation (MSD) of the mean field $h(n)$ 
($\delta h^2 = {\langle h^2 \rangle}-{{\langle h \rangle}}^2$ with $\langle \cdots \rangle$ 
here meaning the long time average) 
should decrease proportionally to $1/N$ by the law of large numbers (LLN) and 
the $h(n)$ distribution must be a gaussian for sufficiently large $N$ 
by the central limit theorem (CLT).  
However there is a certain threshold in $N$ (depending on both $a$ and $\varepsilon$) 
above which MSD ceases to decrease even though the distribution remains gaussian;
CLT holds but not LLN \cite{kb1}. 
This reflects some hidden coherence between the maps in evolution.  
In fact LLN is restored when a tiny noise term is introduced in each map independently \cite{pikovskyb,kb2}.

(iii) The violation of LLN reflects that the map probability distribution $\rho(x)$ 
depends on time. Indeed a noise intensity analysis of ensembles successfully proves LLN \cite{pikovskyb}.
 If LLN should hold in the time average, $\rho(x)$ would have to be a fixed point distribution 
of the Frobenius-Perron(FP) evolution equation \cite{Ruelle}. 
It has been argued that the fixed point distribution may be unstable due to the periodic windows 
of the logistic map \cite{kb2}, though this point is controversial. For instance, on tent
maps, the same instability occurs but no periodic windows are present \cite{ershov}.
The coherence manifests itself in the mutual 
information \cite{kb1}. On the other hand the temporal correlation function 
similar to the Edwards-Anderson order parameter for the spin glass \cite{spinglass}  
decays to zero exponentially. Thus it may not be due to freezing between two elements \cite{kb1}.

The hidden coherence was found in the statistical analysis of 
the mean field fluctuation \cite{kb1}.
But there has been no report of an extensive statistical analysis which covers the whole 
turbulent regime as well as a wide range of the system size.  And once we have done it, 
we are faced with a bizarre feature of the turbulent regime; 
the hidden coherence is one thing but there occur also drastic global periodicity manifestations. 
The above lists are correct but need reservations.

As for (i) there is a need for a careful reservation on the coupling values.  
We show below that, when the coupling $\varepsilon$ takes small 
but tuned values for a given $a$, the maps again -  like in the ordered phase - may split into 
a few bound clusters in periodic motion. 
The most striking manifestation of periodicity in this form is 
the states of almost equally populated clusters mutually oscillating in the same period 
with the number of clusters.
 We call this type of a periodicity manifestation as  
{\it a maximally symmetric cluster attractor} (MSCA) and present below the tuning condition for it. 
We label such a cluster attractor by the periodicity and the number of the clusters.
For instance, we call the outstanding period three symmetric cluster state as $p3c3$ MSCA.
There also occurs bifurcated $p3c3$ MSCA. The MSD of the $h(n)$ distribution is very small 
in the MSCA or its bifurcated state because of the good population symmetry among the clusters. 
At slightly larger $\varepsilon$, we observe that the number of clusters decreases while the orbits are approximately kept. The cluster attractor of this type ($p > c$) leads to large MSD, which
is independent from $N$.

As for (ii) we show below not only LLN but even CLT is violated in almost all regions 
in the turbulent regime.  We pin down the very limited regions in the turbulent regime 
where the CLT holds with violated LLN; only there the term hidden coherence may be used.

As for (iii) the decay exponent of the temporal correlator of the mean field fluctuations 
gradually decreases with the deviation of the coupling from the tuned value. 
Accordingly the $h(n)$ distribution successively changes its shape 
from the highest rank sharp delta-peaks down to the MSD-enhanced Gaussian 
distribution---the hidden coherence.  
This indicates that the hidden coherence at the MSD valley may be the most modest periodicity 
remnant, being elusive due to high mixing.

The GCML can be defined in a one line equation but its turbulent regime 
challenges us with so many faces ranging from a manifest periodicity to the hidden coherence. 
We consider that it is important to explore the systematics of periodicity manifestations 
by an extensive statistical survey and present a sorted list of phenomenological 
observations.  
Below we firstly devote ourselves to this task.  

%%%%%%%%%%%%%%%%%%%%%%%%%%%%%%%%%%%%%
\section{PHENOMENOLOGY OF THE TURBULENT REGIME}

%%%%%%%%%%%%%%%%%%%%%%%%%%
\subsection{Systematics in the mean field fluctuations}
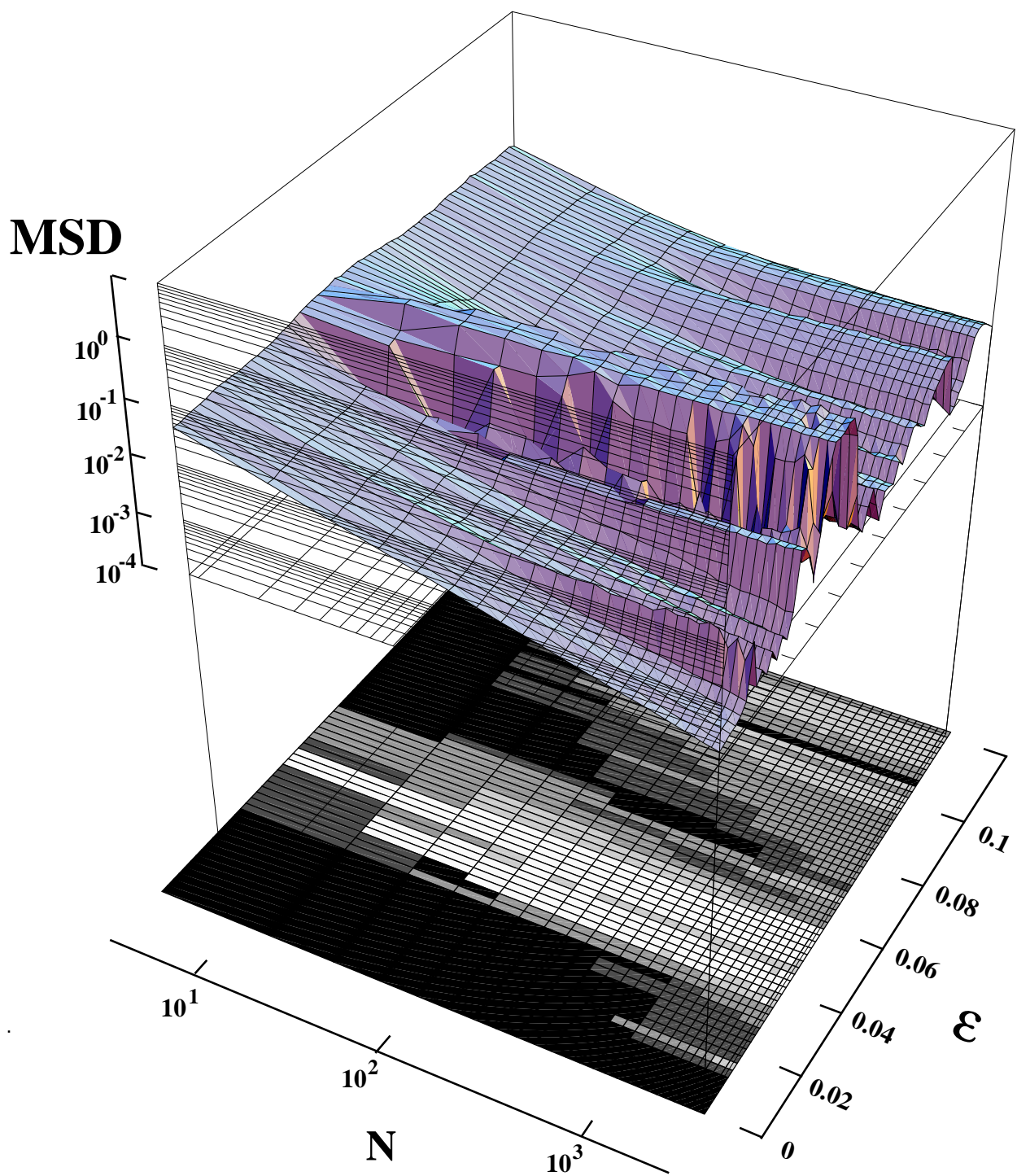
\begin{figure}
\caption{
The MSD surface of mean field fluctuation (top) and 
the gray-scale density plot for the rank of distributions (bottom)  
on the $\varepsilon-N$ grid. $a=1.90$ and 
the inclement $\Delta \varepsilon =2 \times 10^{-3}$.
The rank varies from zero(black) to rank four(white).  
\label{msdsurface}
}
\end{figure}

We start with an analysis of the distribution of the mean field fluctuation in time.
In Fig.~\ref{msdsurface} we show its MSD at $a=1.90$ 
as a surface over the $N- \varepsilon$ plane, which overlays a density plot of the periodicity-rank of the $h(n)$ distribution. 
In order to set sufficiently fine grids for the surface, the system size is limited in the range 
$N < 4\times10^3$. For a wider range of $N$, we show in Fig.~\ref{thermolimit}
the sections of the MSD surface at $N$ in powers of ten up to $10^6$.

%%%%%%%%%%%%%%%%%%%%%%%%%%%%%%%%%%%%%
\subsubsection{Peak-valley structure of the MSD surface}
First let us discuss the MSD surface and its sections.
The linear edge of the surface at $\varepsilon=0$ is of course due to LLN.
For a non-zero but very small $\varepsilon$ $(\lesssim 0.01)$, the LLN still holds to a good 
approximation but otherwise we can clearly see that 
the surface has many peaks along the $N-$axis---the violation of LLN in the time-series\footnote{%--
This does not imply the real violation of the LLN in the ensemble average \cite{pikovskyb}. 
The violation of LLN means here that there is a larger fluctuation in the time series of $h(n)$ 
than expected by LLN.  We are interested in detecting the coherence among the evolving 
elements by the enhanced MSD.}.
%------------------------------------------------------------------
There is a prominent peak at $\varepsilon \approx 0.040-0.050$---an extreme violation of LLN---and in 
this $\varepsilon$ range the MSD is in excess even for $N \approx 10^2$.  
In front of the peak there is a deep valley around $\varepsilon \approx 0.035$.  
We show below these peak and valley are respectively induced by  $p3c2$ cluster attractor and $p3c3$ MSCA 
(and its bifurcated state).  Apart from these, the MSD surface systematically shows successive peak-valley 
structure at large $N$.
We can see clearly in Fig.~\ref{thermolimit} how this structure is formed with the increase of $N$.  
At MSD peaks the LLN violation starts as early as $N=10^2-10^3$ while in the valleys 
one has to wait until $N=10^3-10^4$ in order to observe it.
(In both cases, it starts prevailing from the larger $\varepsilon$ side.) 
This two-fold occurrence of the LLN violation leads to the successive peak-valley structure 
around $N \approx 10^3$, which becomes outstanding at $N \approx 10^4$. 
Beyond that, up to the largest analyzed $N$ (=$10^6$), the MSD is independent from $N$
except for $\varepsilon \lesssim 0.005$ where the maps still follow LLN approximately.

%%%%%%%%%%%%%%%%%%%%%%%%%%
\subsubsection{Mean field distributions}
Now let us discuss the mean field distributions.
In the rank density plot --- the bottom panel of Fig.~\ref{msdsurface} --- the distribution is assigned 
a rank as follows.  
\begin{itemize}
\item[0:~]
The distribution is gaussian\footnote{%--------------------------------
The $h(n)$ distribution cannot be a precise gaussian as is limited in $[-x_L,x_L]$.  
When we discuss whether it is gaussian or not, we concern whether the essence of 
CLT, that the convolution of independent random distributions peaks like gaussian, is 
in action or not.
 }. %------------------------------------------------------
The MSD is the same within 20 percent error with that at $\varepsilon=0$ with common $a$ and $N$. 
\item[1:~]
Gaussian but with a sizably enhanced MSD (the hidden coherence). 
The MSD can be even factor of ten larger than the MSD by LLN. 
\item[2:~]
A singly peaked distorted gaussian, or a trapezoidal distribution. 
\item[3:~]
Either it has a few sharp peaks on top of a broad band, or it is an apparent overlapping gaussian distribution. 
It manifestly shows the periodic motions of the maps. 
\item[4:~]
The distribution consists of a few sharp peaks only. 
\end{itemize}
At rank zero the $h(n)$ distribution obeys both LLN and CLT and the maps may be thought as independent random 
numbers with a common probability distribution.  
Oppositely at rank four the maps are in periodic motion and so is the mean field.
The ranks are organized in a way that the periodicity of the elements becomes more manifest 
with an increase of the rank.  
The MSD surface and the rank density plot both together reveal a simple rule:
{\it The MSD is high wherever the rank is high and vice versa.
The rank distribution plot is almost a contour plot of the MSD surface}\footnote{%-----------
The rank assignment to each of thousands of distributions was a painful task.
It was thrilling that independently determined two diagrams turned out in perfect match.}. 
%------------------------------------------------------------------------------------------------------------

%%%%%%%%%%%%%%%%%%%%%%%%%%
\subsubsection{The regularity in the MSD enhancement}
The above rule persists for larger $N$ too. 
In Fig.~\ref{thermolimit}, we show for $N=10^6$ the $h(n)$ distributions at MSD peaks in the upper 
small boxes and at valleys in the lower.  We find\footnote{%--------------------
These two rules actually hold for $N=10^4$ up to the largest $N(=10^6)$ of our analysis. 
See below for further discussion on the $N$ dependence of the $h(n)$ distribution.}:
%---------------------------------------------------------
\begin{itemize}
\item[(i)] At any MSD peak, the rank is always high---rank three.
This succinctly tells that the high MSD is induced by the maps evolving in quasi-periodic motion 
at the peak $\varepsilon$ values.  
\item[(ii)] On the other hand, at any MSD valleys, the rank is one (the MSD-enhanced gaussian)
and reflects the hidden coherence.
\end{itemize}
In short: {\it the MSD peaks at large $N$ come from the quasi-periodic motion of the element maps
and the hidden coherence is restricted to the MSD valley at large $N$}. 

We should add that the most prominent MSD peak and the deepest valley at the front of it
are two extremes. At the former ($0.040 < \varepsilon < 0.050$) the distribution is 
either rank three or even four and the MSD peak starts even at small $N$. 
The rank four distribution exhibits a periodic coherent motion of maps. 
We show below that it is due to the formation of $p3c2$ cluster attractor; the lack 
of one cluster leads to a high MSD.  
At the latter, the distribution is also rank four but the MSD suppression is realized 
by the symmetrically populated $p3c3$ MSCA. 
We will further investigate the periodicity manifestation in general shortly below introducing other 
phenomenological means too.

%%%%%%%%%%%%%%%%%%%%%%%%%%
\subsubsection{The hidden coherence revisited}
The coherence, as is observed by the violation of LLN, occurs at any $\varepsilon$ value in the range 
$0.005-0.12$ except that the onset of the violation is earlier at MSD peaks. [See Sec. III.A.1].  
But the {\it hidden} coherence implies more; the MSD must be enhanced but the mean field distribution 
must remain Gaussian --- the rank must be one. 

To pin down the regions of the hidden coherence on the $\varepsilon-N$ plane, let us investigate the 
change of the $h(n)$ distribution with the increase of $N$.
Fig.~\ref{Ndepfig} exhibits a typical case; the $\varepsilon=0.0682$ corresponding to one of MSD 
peaks at $a=1.90$.
Just when the LLN violation starts at $N=10^2 \sim 10^3$, the rank becomes one. 
But notably, for $N$ beyond $10^3$, the rank soon becomes two and simultaneously 
the MSD peak-valley structure turns out.
For $N \gtrsim 10^4$, the rank becomes three and the peak-valley structure becomes remarkable. 
The regions of the hidden coherence are thus restricted to a very small part. 
Excepting the transitive region $N \approx 10^2 -10^3$, 
it has to be only MSD valleys for the $h(n)$ distribution to remain Gaussian 
and further $N \gtrsim 10^4$ for the MSD to be enhanced.  
(The deepest MSD valley must be also excepted since we observe the apparent periodic 
motion of $p3c3$ MSCA. ) 

%%%%%%%%%%%%%%%%%%%%%%%%%%
\subsection{The periodicity manifestation in the turbulent regime}

The MSD peak-valley structure reflects a periodicity manifestation in the turbulent regime 
at various strength depending on the value of $\varepsilon$. 
Let us substantiate this issue by the following quantities; 
(i)  the distribution of maps and their mean field, 
(ii)  map-orbits, 
(iii) the temporal correlator of maps\footnote{%------------------------
$  C(t)=\Biggl\langle  
          \tilde{\mbox{\boldmath $x$}}(n+t)\cdot \tilde{\mbox{\boldmath $x$}}(n)/
          | \tilde{\mbox{\boldmath $x$}}(n+t)| | \tilde{\mbox{\boldmath $x$}}(n)|    
            \Biggr\rangle$ with the relative vector 
$\tilde{\mbox{\boldmath $x$}}(n)\equiv (x_1(n) -h_n, \cdots, x_N(n)-h_n)$.
The average $\langle \cdots \rangle$ is taken over $n$ for the last 1000 steps.  }%---------------------------------------------------
~and 
(iv) the return map of $h(n)$.
In Fig.~\ref{dynamicsvariation}, the $a$ is set at $1.90$ and above quantities are listed 
in a row for each typical iteration at characteristic $\varepsilon$.
The lattice size is fixed at $N=10^3$ in order to shed light more on the predominant period three window 
than the other windows. 

%%%%%%%%%%%%%%%%%%%%%%%%%%
\subsubsection{The p3c3 MSCA: the event at $\varepsilon=0.036$}  
Let us first investigate the region of the deepest MSD valley. 
In the map-orbits we find clearly three lines showing the period three motion of maps in three clusters
and the map distribution shows three delta peaks.
The mean field (the black circle) is almost constant due to the high population symmetry 
and accordingly the $h(n)$ return map shows almost degenerate three points.
The temporal correlator oscillates in period three. 
All exhibit the formation of $p3c3$ MSCA.

There is a slight subtlety that the state is actually bifurcated --- six clusters of maps 
with almost equal populations in the bifurcated period three motion. This is seen by the tiny split in the 
orbits near zero\footnote{%-------------------------------------
The six orbit points consist of three doublets of points and the two points in a doublet are 
very close each other. We have checked this numerically but the map distribution with 
the bin size $5 \times 10^{-3}$ shows only three delta peaks. }.  %------------ 
For $a=1.90$, always the bifurcated $p3c3$ MSCA is formed at $\varepsilon \approx 0.035$, 
while at slightly higher $\varepsilon$ ($\approx 0.037-0.041$) the final state is 
either a genuine $p3c3$ MSCA (90 percent) 
or unstable period three clusters with high rate mixing (10 percent) 
depending on the initial configurations. 
We come back to this point in the stability analysis section below.

%%%%%%%%%%%%%%%%%%%%%%%%%%
\subsubsection{The p3c2 cluster attractor state: the event at $\varepsilon=0.042$}
At the nearby stronger coupling ($\varepsilon \approx 0.041-0.051$), the maps 
almost always split into two clusters with the population ratio approximately $2:1$ 
and the two clusters oscillate mutually in period three. 
The map-orbits sampled at $\varepsilon=0.042$ clearly exhibit this $p3c2$ state.
The mean field oscillates in period three with a large amplitude due to the lack of one of the MSCA
and hence leads to a prominent MSD enhancement. See also the largely separated three points
in the $h(n)$ return plot as well as the temporal correlator in period three motion.
Note that this high MSD is independent from the number of maps $N$ --- a way of violating
the LLN ---  simply because the large $N$ GCML dynamics is reduced to that of two clusters. 
The MSD is solely determined by the scale given by cluster orbits and the population {\it ratios}. 
As a check let us try an estimate of the MSD. 
For the population ratio $ \theta_1 : \theta_2$ it is given by
\begin{equation}
(\delta h)^2=(S/3)(    \theta_1^2+ \theta_2^2 - 1/3 )   -       
                                (2T/3) (   \theta_1^2+ \theta_2^2  - \theta_1  \theta_2), 
~\theta_1+\theta_2=1
\label{p3c2msd}
\end{equation}
with $S=\sum x_k$ and $T=\sum x_k x_{k+1}$.  
Let us take as approximations $\theta_1 : \theta_2=2:1$ and the orbit points $x_k$ 
at the tangent bifurcation point\footnote{%-------------------------------------------------------
$a=7/4$ and $x_k=2/21+8/(3\sqrt{7})$$\cos{( (\theta+2k\pi )/3 ) }$, $k=0,1,2$ 
with $\theta=\tan^{-1}(3\sqrt{3})$ for the stable set. 
Numerically ($0.9983,-0.7440,0.03140$).}.  
%--------------------------------------------------------------------------------------------------------
Then we obtain $(\delta h)^2=2^5/(3^3 7)\approx0.169$ in good agreement 
with the observed value $0.16 \pm 0.01$. 

%%%%%%%%%%%%%%%%%%%%%%%%%%%%%%%%%%%%%
\subsubsection{The peripheral point to the $p3c3$ MSCA: two events at $\varepsilon=0.032$}
Here we have to account for the first transient behavior of the maps.  
In the event (A), the maps drop into $p3c3$ cluster attractor 
after a long iteration (at $ n \approx 8\times 10^4$), while in (B), they remain 
in a few unstable clusters in mutual period three motion until the last.
The event (A) is essentially the same with the $p3c3$ MSCA event. We should only note 
that the broad lower band in the $h(n)$ distribution is an artifact of the first transient motion of maps.
In (B), the clusters are unstable and there is a mixing of maps between the clusters; hence we 
can see only three clouds in the $h(n)$ return map. 
But the mixing rate is not so high 
as we can see from a gradual exponential decay\footnote{%-----------------------
We quote by $\tau$ the number of steps in which the correlator decreases to $10^{-3}$
as an estimate of the mixing rate.} of the correlator with $\tau \approx 140$, 
which clearly shows {\it a damped oscillation in period three}.
%-------------------------------------------------------------------------------------------------------- 

%%%%%%%%%%%%%%%%%%%%%%%%%%%%%%%%%%%%%
\subsubsection{The variation of dynamics with $\varepsilon$}

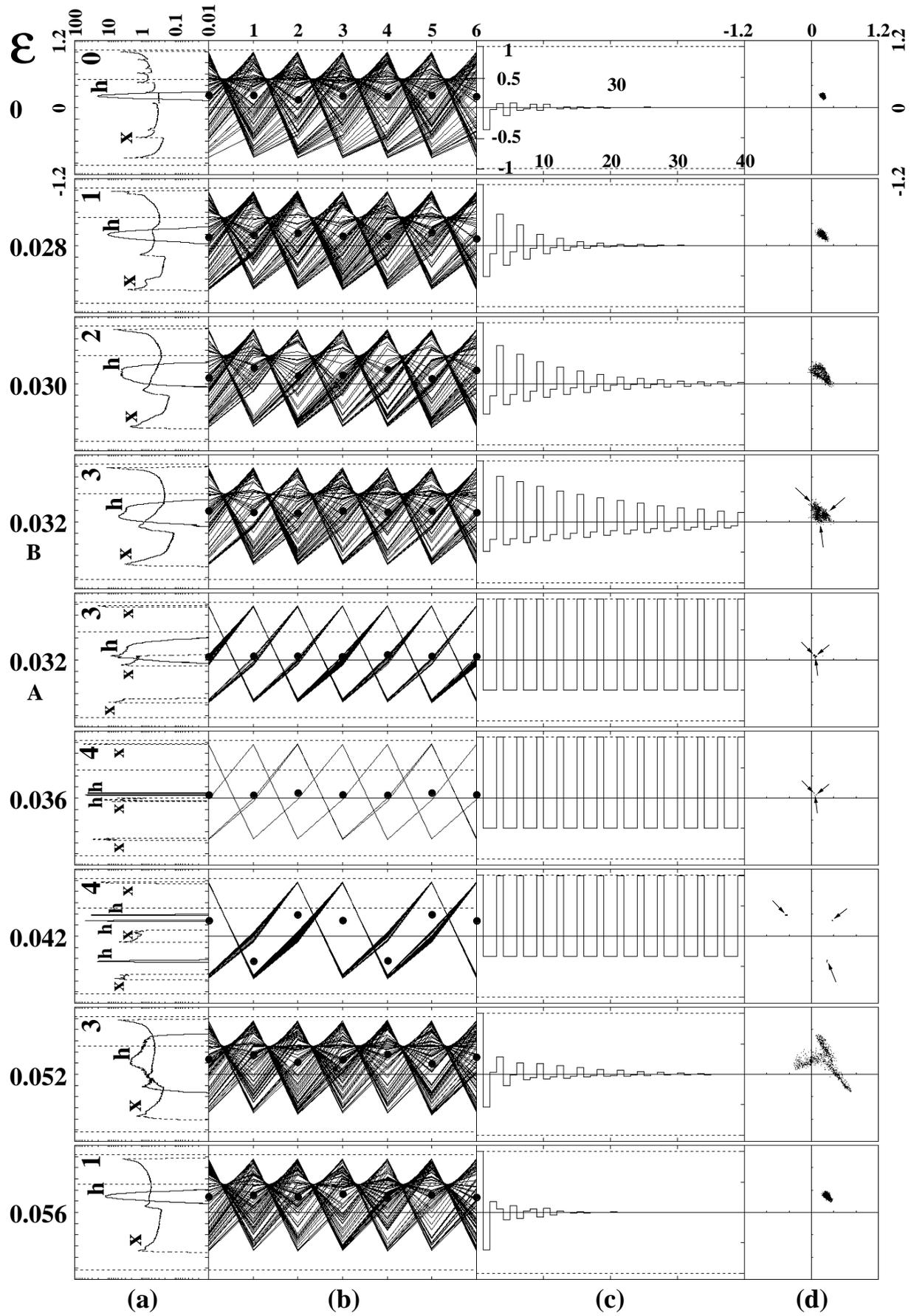
\begin{figure}
\caption{
The variation of GCML dynamics with $\varepsilon$ through the prominent MSD peak region. 
$N=10^3$ GCML with $a=1.90$ and total iteration steps are $10^5$ for each run.
(a) The mean field distribution (marked as $h$) sampled discarding the first $10^4$ transient steps
and the map distribution ($x$) averaged over the last $2 \times 10^3$ steps. 
(b) Clustering pattern. The lines are orbits of randomly selected $10^2$ maps for 
the last seven steps and the black circle is the mean field. 
(c) Temporal correlator between the two relative-coordinate vectors of maps. 
Averaged over the last $10^3$ steps. 
$\tau \approx 0, 30, 70, 140, \infty, \infty, \infty, 50, 0$ for $\varepsilon=0,\cdots,0.056$.
(d) The return maps for the same steps as (c). The arrows indicate the period three 
clusters.
\label{dynamicsvariation}
}
\end{figure}

Let us have a bird's eye view of Fig.~\ref{dynamicsvariation}.
From the row $\varepsilon=0$ to $\varepsilon=0.042$ is the path from randomness to periodicity. 
At $\varepsilon=0$ the maps evolve freely in pure randomness. 
We observe in the map distribution many sharp peaks with fractal structure.
These reflect unstable fixed points of a single map. But the $h(n)$ distribution --- the convolution 
of the map distribution --- is gaussian due to CLT.  It is sharp due to LLN. 
The maps evolve randomly in a simple logistic pattern and the correlator 
decays almost instantly.
With increasing coupling $\varepsilon$, the coherence between maps is increased.
The correlator reveals the precursor of the period three cluster attractor by its $p=3$ oscillation
and becomes prolonged. The map distribution turns into three broad bands losing sub-peaks 
and becomes finally sharp three delta peaks. Because of the increased coherence, 
the $h(n)$ distribution retains the orbits structure even after the convolution and the 
rank of the $h(n)$ distribution is gradually increased. Finally the rank-four distribution 
appears in the period three region.

In the the period three region, we first observe the formation of the $p3c3$ MSCA 
and at slightly higher $\varepsilon$ the $p3c2$ cluster attractor.
This region continues up to $\varepsilon \approx 0.050$.

Beyond this, everything proceeds reversely till $\varepsilon \approx 0.058$.
The rank gradually decreases and the correlator 
gets shortened. Fig.~\ref{dynamicsvariation} ends at this position.
At larger $\varepsilon$, the MSD shows small peaks and valleys
in the range $\varepsilon \approx 0.06 - 0.1$.  Above $\varepsilon \approx 0.10$ 
the correlator catches the precursor of the ordered two clustered regime. 
The path to the periodicity is repeated and eventually the period two regime starts 
around $\varepsilon \approx 0.2$.

This is the bird's eye view of the turbulent regime at $a=1.90$ and $N=10^3$.
For larger $N(\gtrsim 10^4)$, the $MSD$ surface shows peaks and valleys more remarkably. 
The bulk of above variation of dynamics with $\varepsilon$ holds also at the local scale --- for 
each couple of nearby peak and valley. 
At the peak $h(n)$-distribution is rank three and, with the change of $\varepsilon$ 
to the nearby valley values, the rank gradually decreases down to one. 
At the higher (lower) nonlinearity $a$, we observe the same dynamics if the coupling is shifted 
to the larger (smaller) side with appropriate amount.
For instance, the period three attractor region $\varepsilon \approx 0.032-0.050$ 
at $a=1.90$ is shifted to $\varepsilon\approx 0.06-0.08$ at $a=2.0$.
This suggests curves of the balance in the $a,\varepsilon$ parameter space. 
But why the period three attractor states are formed at the particular $\varepsilon$ region? 
Aren't there any other cluster attractors with different periodicities? 
Our next task is to answer these questions deriving the curves of the balance analytically.
%%%%%%%%%%%%%%%%%foliation%%%%%%%%%%%%%%%%%%%%

\section{An analytic  approach} 

\subsection{The tuning condition and period three clusters} 

Let us consider an idealized (exact) MSCA. It  is a state of GCML under three conditions. 
(i) The $N$ maps of GCML split into $c$ clusters with an exact population symmetry,
(ii) the synchronization of maps is perfect so that there is no variance of map positions in
each of the clusters,
and (iii) the clusters mutually oscillate around $p=c$ orbit points.
Using this idealized state as a key,  we derive below the tuning condition for the MSCA formation.   
 For brevity we explain our approach 
with respect to the $p3c3$ MSCA in detail but everything below also goes through for 
the other MSCA with higher periodicity. 
In a $p3c3$ MSCA three clusters A,  B, C move cyclically round three fixed positions 
$X_1, X_2,  X_3$.
Such a  system of orbit points exists as a triple intersection point of three surfaces given by 
\begin{eqnarray}
   \Sigma_i:~~X_{i}=1-a \left[X_{k}^2+\frac{ \varepsilon}{3} 
(X_{j}^2+X_{k}^2-2X_{i}^2) \right], ~~  (i,j,k) 
\in \left\{ (1,2,3),(2,3,1),(3,1,2) \right\}.  \label{cyclicequation}
\end{eqnarray}
At $a=1.90,\varepsilon=0.040$, for instance, we have two solutions 
$(0.96301, -0.00499,  -0.72851)$ and $(0.95521, 0.07076,  -0.69993)$; 
the former is stable and the latter is unstable.
In such an exact MSCA, the meanfield $h(n)$ is a time-independent constant:
\begin{eqnarray}
   h(n) \equiv \frac{1}{N} \sum_{i=1}^N f(x_i(n))= \frac{1}{3} \sum_{I=A,B,C} f(X_I(n))=
 \frac{1}{3} \sum_{i=1}^{3} f(X_i)
= \frac{1}{3} \sum_{i=1}^{3} X_i\equiv h^* ,
\end{eqnarray}
where $X_I(n)$ denotes the coordinate of the cluster $I$ at time $n$
and the last equality follows from (\ref{meanfield}) or (\ref{cyclicequation}).  
Therefore, if MSCA is produced, the GCML evolution equation ($\ref{global1}$) becomes 
\begin{eqnarray}
    x_i(n+1)&=&(1-\varepsilon)f_a(x_i(n))+\varepsilon h^* ~~~(i=1, \cdots, N), \label{consthevolution}\\
    f_a(x)    &=& 1 -a x^2,  \nonumber
\end{eqnarray} 
where the time-dependent term $h(n)$ is replaced by a constant $h^*$.
Every one of the maps evolves by a common equation at each step in ($\ref{global1}$) 
and further by a unique constant equation in (\ref{consthevolution}).
As is noted by Perez and Cerdeira \cite{PerezCerdeira} some years ago, we can 
cast this unique equation to a standard logistic map with a reduced
nonlinear parameter $b$
\begin{eqnarray}
    y_i(n+1)=1- b \left(y_i(n)\right)^2  ~~~(i=1, \cdots, N ).   \label{logisticb}
\label{yevolution}
\end{eqnarray} 
by a linear scale transformation  
\begin{eqnarray}
    y_i(n) = (1-\varepsilon+\varepsilon h^*)^{-1} x_i(n)  \label{lineartransformation},
\end{eqnarray}
and the reduction rate of the nonlinearity parameter is given by
\begin{eqnarray}
   r \equiv \frac{b}{a} = (1- \varepsilon)  \bigl( 1 - \varepsilon (1- h^*) \bigr)  \label{rconstraint}.
\end{eqnarray}
At MSCA, the mean field $h^*$ is constant, so the reduction factor $r$ is also constant. 
If the clusters of  MSCA oscillate in period three, so do the maps $y_i(n)$ --- the two solutions 
$(x_1,x_2,x_3)^{(\nu)},~ (\nu=1,2)$ of the cyclic equation (\ref{cyclicequation}) agree with the two sets of 
period three orbit points $(y(n),y(n+1),y(n+2))^{(\nu)},~(\nu=1,2)$ of the logistic map (\ref{logisticb}) modulo 
the scale factor in (\ref{lineartransformation}).
The reduction factor $r$ must reduce the high nonlinearity $a$ of GCML down to the $b$ in the period three 
window. 
It starts at $b_{\text{th}} \equiv 7/4$ by the tangent bifurcation and, after sequential bifurcations 
and windows in the window,  it closes at $b = 1.79035$ by the crisis.    
The range of the period three window $b=1.75-1.79035$ requires a reduction 
factor $r$  in the range $0.942 - 0.921$ for $a=1.90$. 
Each $r$ within this range gives a constraint curve\footnote{%----------------------------------------
It is possible to transform formally  (\ref{global1}) to a standard form 
at each step but then the reduction factor $r$ may fluctuate step by step. 
Then it does not single out a line.} 
%--------------------------------------------------------------------------------------------------------------------- 
on $\varepsilon, h^*$ plane via  (\ref{rconstraint}).

There is another constraint from self-consistency; the average value $y^*$ of the transformed maps must also 
obey (\ref{lineartransformation}) so that 
\begin{eqnarray}
    y^* \equiv \frac{1}{3} \sum_{i=1}^3 y_i = (1-\varepsilon+\varepsilon h^*)^{-1} h^*.  \label{yconstaraint}
\end{eqnarray} 
Here $y^*$ is a function of the nonlinear parameter $b$---it is simply an equal weight average 
of the period three stable orbits of the single logistic map ($\ref{yevolution}$) 
and can be estimated solely by 
the property of the logistic map without any recourse to the GCML evolution equation. 
At a given $y^*$ this again gives a constraint curve on $\varepsilon, h^*$ plane. Let us work out the $\varepsilon$
at the intersection of the two curves.
By eliminating $h^*$ from (\ref{rconstraint}) and (\ref{yconstaraint}) we obtain
\begin{eqnarray}
    \varepsilon = 1 - \frac{r y^*}{2}  - \sqrt{
 r (1- y^* ) +  \left(\frac{ry^*}{2}\right)^2 }. \label{tuningcondition}
\end{eqnarray}  
and  both $r=b/a$ and $y^*$  in the right hand side are determined by $b$.
This is the {\it tuning condition}. This predicts the {\it necessary value} of the coupling $\varepsilon$ for 
the GCML at a given $a$ to form MSCA due to the periodic attractor of the single logistic map with $b$.
   The function $y^*(b)$ is a well-known square-well.  
$y^* \approx 0.284$ at $b=1.735$ slightly below the tangent bifurcation point $b_{\text{th}}$, 
and it drops sharply ($y^* - y^*_{\text{th}} \propto \sqrt{b_{th}-b} $) at $b_{\text{th}}$. 
From matching of the coefficients in 
$(f_{b})^3(y) -y = b^6 (f_{b}(y) -y)\left(\Pi_{i=1}^3(y-y_i)\right)^2$,
 we obtain $ y^*_{\text{th}}=1/(3 \cdot 2 b) = 2/21=0.095 $.
Similarly $y^*(b) =(1-\sqrt{4b-7})/6b$ up to the first bifurcation point $ b=1.769$.
Then, $y^*(b)$ varies smoothly\footnote{%
The largest rapid variation is the tiny anti square 
well ($\Delta y^* \approx 0.01)$ due to the $3 \times 3$ window at $b=1.7858-1.7865$}
%----------------------------------------------------------
around $0.08$ until the end of the window ($b=1.7903$)
and finally increases sharply ($y^* \approx 0.18$ at  $b \approx 1.793$).     
This $y^*(b)$ put into (\ref{tuningcondition}) gives the following 
estimates of $\varepsilon$ for $a=1.90$; 
\begin{eqnarray}
   A: ~~   \varepsilon &=& 0.0514   \mbox{~~~at~~~} 
         (b,  y^*)=(1.735, ~0.284),   ~r=0.913  \nonumber \\
   B: ~~   \varepsilon &=& 0.0422   \mbox{~~~at~~~} 
         (b, y^*)=(1.750, ~0.095),   ~r=0.921    \nonumber \\
   C: ~~   \varepsilon &=& 0.0363   \mbox{~~~at~~~} 
         (b, y^*)=(1.769, ~0.069),   ~r=0.931    \label{estimates}  \\
   D: ~~  \varepsilon  &=& 0.0305   \mbox{~~~at~~~} 
         (b, y^*)=(1.790, ~0.080),   ~r=0.942    \nonumber  
\end{eqnarray}
% A: e=0.0523201, r=0.91, (b=1.729)  
% A: e=0.0514397, r=0.91315, (b=1.7350)  
% B: e=0.0422164, r=0.921, y*=0.095  (b=1.75)
% period3 window starts at b=1.75,    (r=0.92105)
% C: e=0.0364547, r=0.93079, y*=0.07 (b=1.76851)
% period6 starts at b=1.76851            (r=0.93079)
% D: e=0.0304678, r=0.94229, y*=0.08   (b=1.79035)
% period3 window ends at b=1.79035 (r=0.94229)
The estimates A, B, C and D are respectively below the threshold, at the threshold, 
at the first bifurcation point in the window, and at the closing point of the window.
Note that the route $\mbox{A}\rightarrow\mbox{D}$ is in the direction of increasing $b$, which in turn
is the direction of decreasing coupling constant $\varepsilon$, since the larger $b$ requires only a smaller nonlinearity reduction.  
We should  stress that the tuning condition (\ref{tuningcondition}) is a necessary condition. 
For the $p3c3$ MSCA to be stable, 
the orbit of the reduced logistic map must be also stable\footnote{%--------------------------------
As we show in the stability section below, the Lyapunov exponents of GCML at the $p3c3$ MSCA
consist of $N-3$-fold degenerate one and three in general non-degenerate ones.
For the $p3c3$ MSCA to be stable, at least the former degenerate exponent must be 
negative, which implies the reduced map orbit must be stable.
That all these exponents are negative at MSCA is shown also below.}. %-----------------------------------
Period three logistic orbit still continues to exist even beyond the first bifurcation point C, 
but it is unstable. Therefore, for an exact $p3c3$ MSCA to be formed, the $\varepsilon$ range 
must be within the estimates C-B, but neither within D-C nor beyond D. 
Similarly an exact bifurcated $p3c3$ MSCA must be formed 
within D-C. Our tuning condition does not guarantee the formation of the MSCA 
but it does limit the $\varepsilon-$range in which the formation is possible.
The observed ranges of the $p3$ cluster attractors are listed in Table \ref{tablep3c3}.
At $a=1.90$, $p3c3$ MSCA is formed in the range $\varepsilon \approx 0.037-0.041$
and its bifurcated state in $\varepsilon  \approx 0.032-0.037$.
The predicted ranges are respectively $\varepsilon = 0.0363-0.0422$ and 
$\varepsilon = 0.0305-0.0363$. 
In both cases, the agreement is remarkable and we see that the formation 
actually occurs at any allowed $\varepsilon$ value.

As for the $p3c2$ cluster state, we need a caution in using the tuning condition.
It is derived under the assumption of the constancy of the mean field. 
Thus, as a matter of principle, it cannot be applied for the asymmetrically populated state.  
However, the $p3c2$ state is formed with a slightly higher coupling $\varepsilon$ 
and the orbits of two clusters are approximately the same with the MSCA orbits. 
Therefore, the $p3c2$ cluster attractor is certainly still under the control of the period three window.
We estimate the range by the extension of the period three window at the higher coupling side
B-A --- the intermittency region. This gives $\varepsilon=0.0422-0.0514$ in good agreement with 
the observed range of the $p3c2$ cluster attractor ($\approx 0.041-0.050$).
 It is interesting to note that the GCML final states at this $\varepsilon$ range actually 
consist of two types depending on the initial condition; the 
$p3c2$ cluster attractor ($ \approx 80 \% $) as well as the unstable period three clusters 
with mixing of maps (the rest).  See Fig.~\ref{lyapunovedep} below.
The estimate by B-A relates intriguingly the intermittency of the element map to 
the GCML phase of co-existent stable and unstable periodic clusters. 
We are aware that we cannot take the success of the estimate for $p>c$ states on the same 
footing with that for the MSCA but at least it gives a good rule of thumb for the $p>c$ state.

\subsection{Foliation of the logistic windows in the turbulent regime} 

What is the case for other $a$ values?
Do the other windows also show up in the expected $\varepsilon$ 
range in the turbulent regime?    
To check these systematically, let us note that the tuning condition defines a one-parameter ($b$) 
family of curves in the model parameter space (the $a,\varepsilon-$plane) of the GCML. 
Each curve is labeled by $b$ and written as a function of the reduction factor $r$ as
\begin{eqnarray}
(a^{(b)}(r),\varepsilon^{(b)}(r))= \left(\frac{b}{r},1 - \frac{r y^*(b)}{2}  - 
\sqrt{r(1- y^*(b))+\left(\frac{ry^*(b)}{2}\right)^2}\right),~~r \le 1. \label{foliationcurves}  
\end{eqnarray}    
It emanates from the point $(a^{(b)}, \varepsilon^{(b)})|_{r=1}=(b,0)$ 
and with the decrease of $r$ it develops in the parameter space in the direction in which both 
$a$ and $\varepsilon$ increase in a certain balance.
If our success above is a general one, all of the GCML with the parameters being 
set at $(a^{(b)}(r), \varepsilon^{(b)} (r))$ along a curve labeled by $b$ should 
be commonly controlled by the same dynamics of the single logistic map at $b$. 

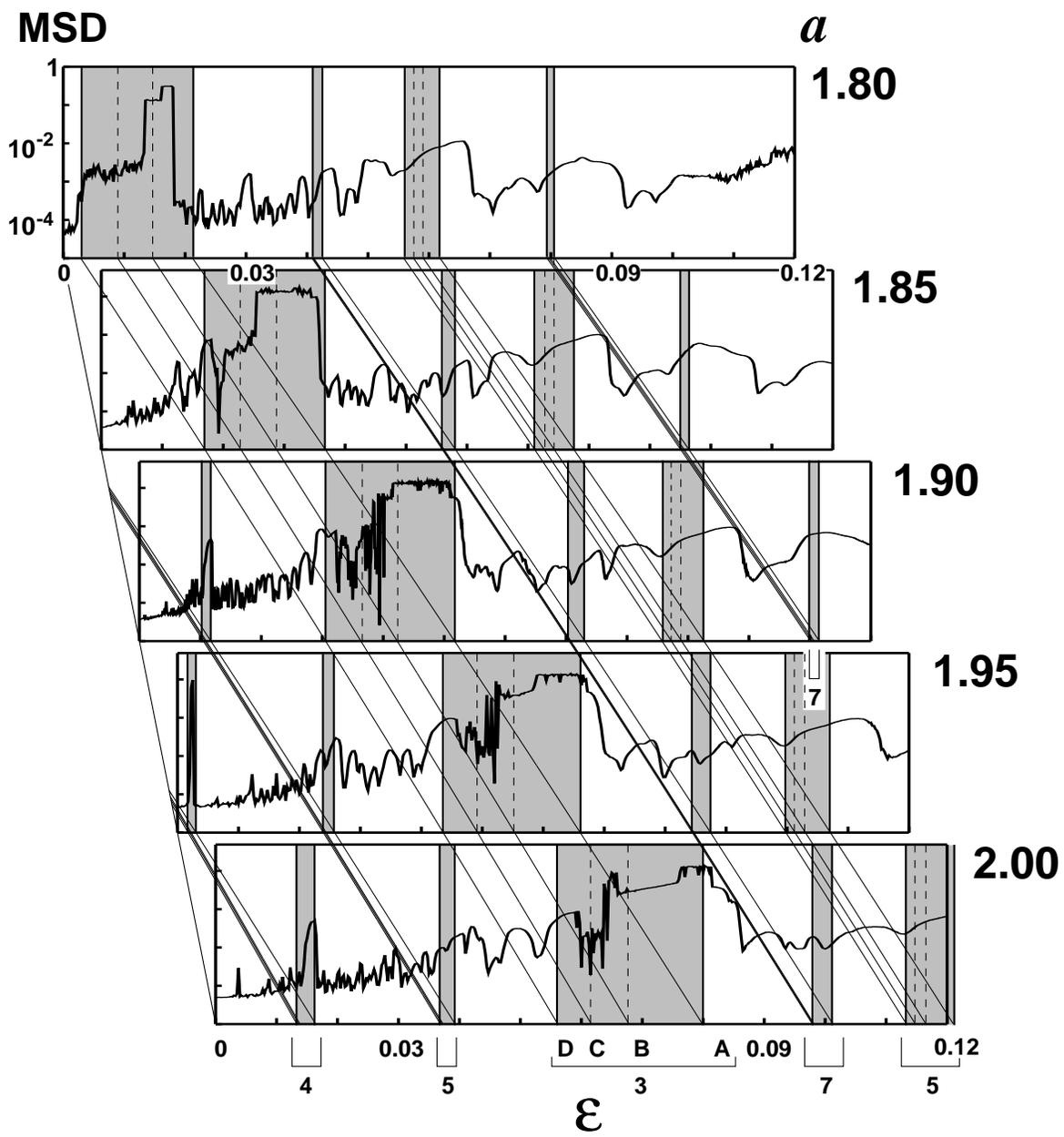
\begin{figure}
\caption{
The MSD of the mean field distribution plotted as a function $\varepsilon$ 
for $N=10^4$ GCML in panels at $a=1.8$(back), $1.85, \cdots, 2$ (front). 
$\varepsilon=0 - 0.12$ with inclement $10^{-4}$. 
The foliation curves predicted for $p=4, 5, 3, 7, 5$ 
windows flow underneath panels and link the respective shaded foliation zones. 
For each zone a MSD-valley at lower $\varepsilon$ and a peak at higher.   
\label{foliation}
}
\end{figure}

In Fig.~\ref{foliation} we find that this is indeed the case.  
Each panel shows the MSD of the $h(n)$ distribution as a function of $\varepsilon$ 
at a given $a$ as well as the expected zones for the manifestation of the outstanding six windows 
in Table.\ref{tableabcd}.  
The curves (\ref{foliationcurves}) are displayed underneath the panels and link the respective zones. 
At each zone, a MSD valley due to MSCA should appear in the lower $\varepsilon$
side and a MSD peak by $p>c$ cluster attractors at the nearby higher $\varepsilon$.
We find that this works with almost no failure in all panels and with respect to all six windows.

The effects of the logistic windows propagate along the curves (\ref{foliationcurves}),
which may be called as {\it foliation curves}.
The curve with the label $b$ links together  those GCML commonly subject 
to the same logistic window dynamics at $b$. 
Accordingly the family of the curves produces the {\it foliation} of the single map dynamics.
The foliation occurs because, under the global interaction, the maps of the GCML form a macroscopically
coherent state. Even though the coupling in the turbulent regime is very small, 
the coherence prevails over the GCML maps if the tuning condition is met.  
A few remarks are in order.

{\noindent \it 1. Desynchronization along the foliation curve}~~
The periodicity manifestation becomes weakened at a higher reduction 
and there is a threshold $r_{\text{th}} \approx 0.95$.
For $r \approx 1$, both MSCA and associated $ p>c $ clusters are formed in tight synchronization.
Towards $r_{\text{th}} \approx 0.95$ the clusters broaden. There is no mixing yet among the clusters  
but the maps move chaotically in each cluster.   
Below $r_{\text{th}}$, we observe only the periodicity remnants --- on one hand the overlapping-gaussian 
$h(n)$-distribution along the curve which had the $p>c$ cluster above $r_{\text{th}}$, 
and on the other hand the MSD-enhanced gaussian (the hidden coherence) along that of MSCA. 
See Fig.~\ref{thermolimit}.

{\noindent  \it 2. Left-right asymmetry of the MSD curves}~~
The MSD curves in Fig.~\ref{foliation} (and \ref{thermolimit}) show an interesting feature --- in each 
panel the smaller $\varepsilon$ region (the left ) has an ample amount of peaks and valleys, 
while the larger only a few broad ones.
As for the single logistic map, on the other hand, there are as many windows in the smaller $b$ as in the larger.  

  This is naturally understood by the difference in the reduction factor $r$ between the zones in a panel.  
In a way, each panel is a screen which displays the windows of the single logistic map by using a macroscopic 
coherent state of GCML. But the panels set at fixed $a$ values are inclined --- a smaller $\varepsilon$ (the left) 
implies less reduction, i.e. $r \approx 1$.  
The $r_{\text{th}}$ divides the panel at $\varepsilon \approx 0.030$ via (\ref{tuningcondition}). 
The left sensitively displays the sharp peak-valley structure induced by cluster attractors. 
The right, on the other hand, can reflect only the accumulation of the periodicity remnants from nearby windows, 
being dominated by the prominent one at its respective zone. 
As a check we set the panels at fixed $r$ values. Then they displayed windows without asymmetry, 
and with a higher sensitivity at $r$ closer to one\cite{shimadakikuchi}. 

%Table.\ref{tableotherpc}
{\noindent \it 3.  Cluster attractors with higher periodicity}~~~Let us search cluster attractors with periodicity 
higher than three. Here we give two samples in Table.\ref{tableotherpc}.

{\it $p=4$ clusters}~~These appear in the left most zone in the $a=1.95$ panel. 
From the window $b$ data in Table.\ref{tableabcd} the necessary 
reduction from $a$ is very small --- $r \approx 0.995$ --- so we expect definite clusters. 
We indeed find the expected sequence of clusters\footnote{%---------------------------------------------
The single cluster cannot be formed. The focusing by averaging does not act there 
and the tiny variance is instantly amplified.
It appears far in the coherent phase ($\varepsilon \gtrsim 0.4$ at $a=1.90$).} %----------------------  
$p=4$, $c=4$(MSCA)$\rightarrow$$3$$\rightarrow 2$ 
in tight synchronization at the right $\varepsilon$.

{\it $p=5$ clusters}~~There are two $p=5$ windows in Table.\ref{tableabcd}.
We choose the one at the lower $b$ and set $b=1.66$ which amounts to $r=0.980$.
Since $r$ is in the mid of one and $r_{\text{th}}$, we expect the clusters are not 
in complete synchronization but yet there is no mixing of maps. 
Indeed the sequence of attractors $p=5, c=5\mbox{(MSCA)} \rightarrow 4 \rightarrow 3$ 
is observed at the expected $\varepsilon$ and it terminates 
before the lowest one ($p5c2$).
%%%%%%%%%%%%%%%%%%%%%%%%%%%%%%%%%%%%%%%%%%%%%%%%%%%%
%--------------------------------------------------------------
\section{Stability of the period three clustered map states} 
%Fig12    e-dependence of linear stability
%Fig13    $p3c3$(MSCA) stability
%Fig14    $p3c2$ stability

%
Here we adopt the Lyapunov analysis.  As one superlative ability, it can be applied to both 
diverging and converging system orbits so that it can detect the possible coexistence 
of multifold finial states depending on the initial configurations. 
We measure the maximum Lyapunov exponent $\lymax$ by a standard method  \cite{nagashima} 
which keeps track of an $N$-dimensional shift vector $\delta {\bf x}(n)$ evolving under 
the non-autonomous linearized equation associated with (\ref{global1});
%----------------------------------------------------------------------------
\begin{eqnarray}
 \delta x_i(n+1)  =  -2a \left\{
                                 \left(  1-\varepsilon +\frac{\varepsilon}{N}  \right) x_i(n) \delta x_i(n) 
                              + \frac{\varepsilon}{N}\sum_{j \ne i} x_j(n) \delta x_j(n) 
                                \right\}.     
\label{linearizedequation}
\end{eqnarray}
The $\lymax$ is the average of the logarithm of the expansion rate of the 
shift vector (with intermediate renormalizations).  For both $\lymax$ and MSD,
we discard the first transient $10^4$ steps  --- generally it takes only 
$10^2 \sim 10^3$ steps for the cluster formation.

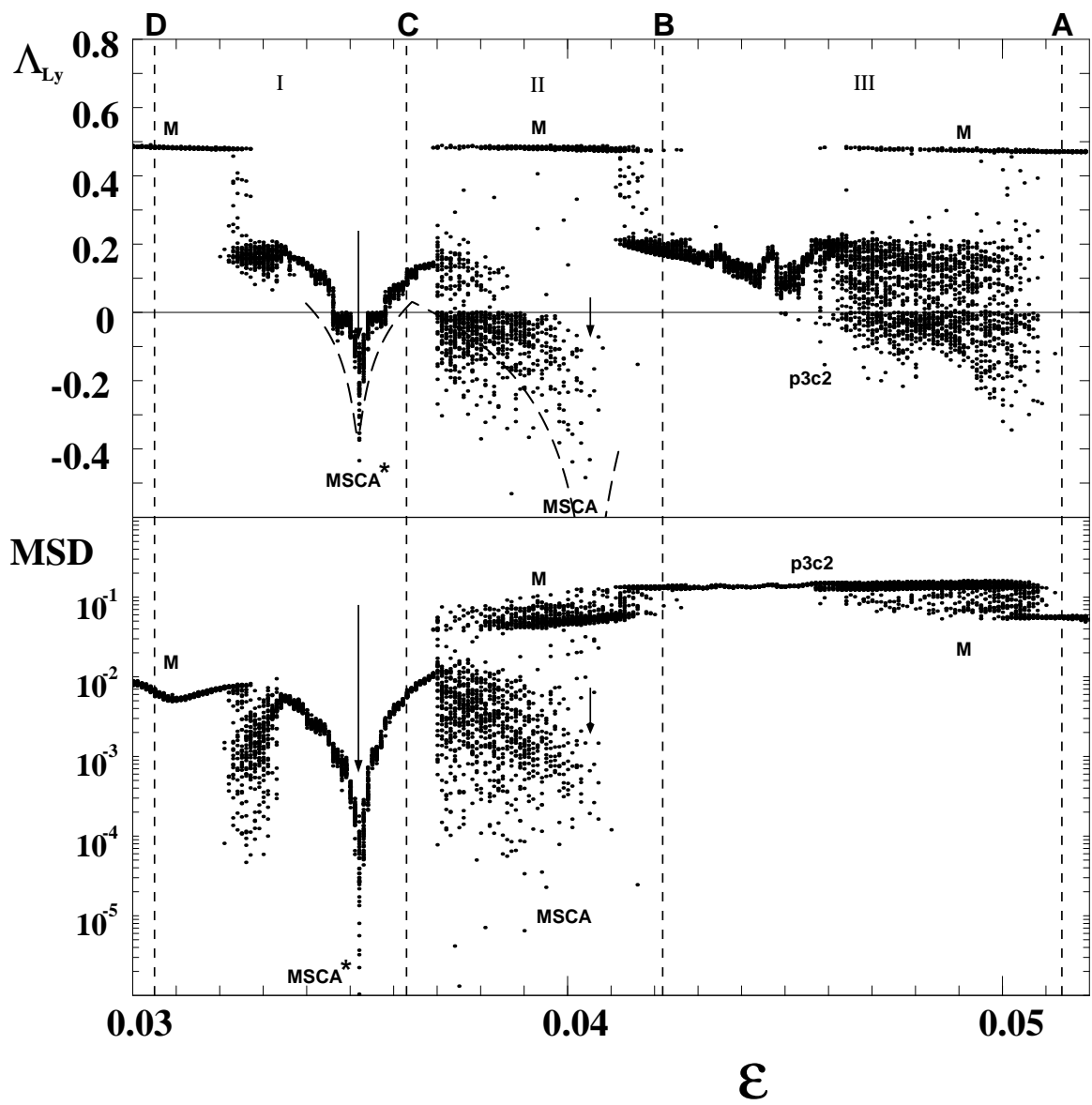
\begin{figure}
\caption{
The Lyapunov exponents (upper) and the MSD (lower) measured for 40 random initial configurations 
at each $\varepsilon$ between $0.0300$ and $0.0520$ with inclement $10^{-4}$. $a=1.90$ and $N=10^6$. 
The events in the seagull structures are in the bifurcated MSCA states, the first lower band events 
the MSCA, and the second ones the p3c2 attractor. These are formed in regions I (D-C), II (C-B) 
and III (B-A) separated by dashed lines just as expected. 
In bands M, the $p=3$ clusters are unstable and maps are mixing among them with 
$\tau \approx 100$.  The dashed line; the predicted $\lymax$ for the exactly symmetric MSCA.
Arrows; the predicted positions of the most linearly stable MSCA states. 
\label{lyapunovedep}
}
\end{figure}

Let us first check the $\varepsilon-$dependence of the stability of attractors.
We choose $N=10^6$, fix $a$ at $1.90$, and vary $\varepsilon$ in the range $0.030-0.052$ 
with the inclement $\Delta \varepsilon=10^{-4}$.  
We show in Fig.~\ref{lyapunovedep} the $\lymax$ in the upper 
and the MSD in the lower\footnote{%----------------------------------------------------------------------
For reference, the CPU time for $N=10^6$ GCML is approximately two minutes for one measurement 
of $\lymax$ ($2^{12}$ steps for precision $10^{-4}$) plus MSD ($10^4$ steps) 
on a modest supercomputer VPP300/6.
The total is $2 ~\mbox{min} \times ~40 \mbox{(initial configurations)} \times 
220 ~(\varepsilon \mbox{-values)} \approx 300 ~\mbox{hours}$.}.%----------------------------------

We observe in the $\lymax$ plot three remarkable structures of low $\lymax$ events.\\
(i) {\it A seagull structure} ($\varepsilon = 0.032-0.037$) with a sharp cusp 
at $\varepsilon=0.0352$ --- all events are bifurcated MSCA  with good population symmetry 
($N_I/N \approx (1 \pm 0.05)/6)$. 
Note that the events form also a seagul in the MSD and the cusp positions agree precisely. 
     {\it The bifurcated MSCA is the more stable if the mean field fluctuation is the less and 
          it is the most linearly stable ($\lymax=-0.38$) with the 
          minimum fluctuation ($\delta h^2 \approx 2 \times10^{-6}$).} \\
(ii) {\it The first low band} ($\varepsilon = 0.037-0.041$) --- The $p3c3$ states. 
The population distributes around the exact MSCA --- $\theta_I \equiv  N_I/N \approx (1 \pm 0.15)/3$.
The events near the lower boundary ($\lymax <0$) are $p3c3$ 
events with good population symmetry and with low MSD. \\
(iii) {\it The second low band} ($\varepsilon = 0.041-0.051$) --- the $p3c2$ cluster attractor.
 The corresponding MSD is, contrary to (ii),  extremely high  because of a lack of one 
cluster to minimize $h(n)$ fluctuation [See Sec. III.B.2]. 

The foliation of the critical points A, $\cdots$,D from the period three window
defines three $\varepsilon$-regions I(D-C), II(C-B), III(B-A) [Eq. (\ref{estimates})]. 
The region I is the allowed region for the formation of the bifurcated MSCA (MSCA${}^*$), 
II the p3c3 MSCA and the $p3c2$ attractor cluster is expected in III. 
As we see clearly in Fig.~\ref{lyapunovedep}, the regions I, II and III respectively embody 
the structures (i), (ii) and (iii) just in agreement with our prediction. 

Let us note a remarkable feature in the events in the two wings of the seagull (i).
Here all events come out with positive $\lymax$ ($\approx 0.1 - 0.2$).  
For a system with low degrees of freedom, the positive $\lymax$ implies chaos. 
But here even with positive $\lymax$, the maps always form bifurcated $p3c3$ state.  
There is actually no contradiction. The global motion of the clusters is periodic, but,
inside each cluster, maps are here evolving randomly with tiny amplitudes 
($\lesssim 10^{-2}$) in sharp contrast against the complete synchronization at the cusp.
The Lyapunov exponent measures the linear stability of the system with respect to 
the small deviation of the element position.  It is sensitive to the microscopic motion of 
the element of the system and hence yields the positive exponent. 
But for a larger deviation, nonlinear terms can become relevant and pull back
the map\footnote{%------------------------------------------------------------------------------------
We have verified this by inputting pulses on randomly selected maps. The analytic formulation
of the nonlinear effect is most wanted for.}. %-------------------------------------------------------------
  This type of map motion --- microscopically chaotic but macroscopically in the 
periodic clusters --- may be called as {\it confined chaos}. 
%--------------------------------------------------------------------------------------------------------------

  We hereafter devote ourselves into the investigation of two outstanding structures, 
namely the bifurcated MSCA seagull and the $p3c2$ cluster attractor.

{\it The linear stability analysis of the bifurcated MSCA}~~  
In order to understand the salient cusp at $\varepsilon=0.0352$ 
in the Lyapunov exponent plot, let us consider the linear stability matrix 
of the GCML.\\
(1) For the configuration of maps in six clusters,  the $N \times N$ linear stability matrix 
of the GCML for evolution of one step can be written as  
\begin{eqnarray}
M_1=(1-\varepsilon)
\left(
\begin{array}{ccc}
X_1 E_1&\cdots&0        \\
0          &\ddots&0        \\
\cdots   &\cdots&\cdots \\
0          &\cdots&X_6 E_6 
\end{array}
\right)
+
\frac{\varepsilon}{N}
\left(
\begin{array}{ccc}
X_1 H_{11}&\cdots&X_6 H_{16}  \\
X_1 H_{21}&\cdots&X_6 H_{26}  \\
\cdots       &\cdots&\cdots  \\
X_1 H_{61}&\cdots&X_6 H_{66}  
\end{array} 
\right) \label{gcmlstabilitymatrix}
\end{eqnarray}
multiplied by an overall factor $-2a$, where 
$X_I (I=1,\cdots,6)$ are the coordinates of the clusters, 
$E_I$ a $N_I \times N_I$ unit matrix, and $H_{IJ}$ is a $N_I \times N_J$ matrix with all elements one.
The $N$ eigenvalues of $M_1$ consist of two sets.
One is a set of $6$ eigenvalues $\lambda^{(I)}= - 2a (1-\varepsilon) X_I,(I=1,\cdots,6)$, 
each $(N_I-1)$-fold degenerate.  
The degenerate eigenvectors of $\lambda^{(I)}$ are of the form  
Col.$({\bf 0}; \cdots ;{\bf 0}; (1, 0,\cdots,0,-1,0,\cdots,0);{\bf 0}; \cdots ;{\bf 0})$,
that is, all column blocks, each for one cluster, are fulfilled by ${\bf 0}$ except for the $I$-th block
which has $1$ as the first element and  $-1$ as one of the other $N_I-1$ elements. 
The eigenvector of  $\lambda^{(I)}$ represents a shift of the system orbits within the $I$-th cluster.  
The other is a set of  (in general non-degenerate) $6$ eigenvalues  
$\lambda_I$, which are the same with the ones of the $6 \times 6$ stability matrix 
$M_{\text{1; red}}$ associated with the cluster evolution 
\begin{equation}
 X_I(n+1) = (1-\varepsilon) f(X_I) + \varepsilon \sum_{J=1}^6 \theta_J f(X_J), ~(I=1,\cdots,6).
\end{equation}
The $M_{\text{1; red}}$ for the cluster dynamics is derived from $M_1$ 
by $E_I \rightarrow 1$  and $H_{IJ} \rightarrow \theta_J$. The eigenvector of $M_1$
subject to $\lambda_I$ is $(\xi_1^{I} {\bf 1};\cdots;\xi_6^{I} {\bf 1})$,
with $(\xi_1^{I}, \cdots, \xi_6^{I} )$ being that of $M_{\text{1; red}}$. \\
(2) 
The stability matrix $M_{p}$ of GCML for the evolution of $p$ steps is given by the chain product  
of $p$ of $M_1$ along the system orbit. 
The eigenvalues of $M_{p}$ again consist of two sets. 
One is the set of $6$ eigenvalues 
$\lambda^{(I)}=[-2a(1-\varepsilon)]^p \prod_{k=1}^p X_I^{k}$, 
each $(N_I-1)$-fold degenerate, and the first set eigenvectors of $M_1$ remain 
the eigenvectors of this set.
The other is the same with the ones of the $M_{p; \text{red}}$---the $p$-th 
iterate of $M_{1; \text{red}}$.
This mechanism holds at any population composition among the GCML clusters.\\
(3)
Now, when the population symmetry among the clusters are exact, all of the maps obey 
a unique quadratic mapping (\ref{consthevolution}) with a constant mean field $h^*$ 
which is equivalent to a standard logistic map (\ref{logisticb}) with a reduced nonlinearity $b$ 
via the scale transformation (\ref{lineartransformation}). For $b$ from the first
to the second bifurcation point in the $p=3$ window($b_6=1.76852915$ to $b_{12}=1.777221618$),
the reduced map $y$ evolves in period six and so do the GCML six clusters.
This is the bifurcated MSCA.  We can write the correspondence as 
\begin{eqnarray}
\begin{array}{ccccccccc}
y_0&\rightarrow & y_1=f_b(y_0)&\rightarrow&\cdots&\rightarrow
&y_5=(f_b)^{5}(y_0)&\rightarrow&y_0=(f_b)^{6}(y_0)\\
\left(\begin{array}{c}X_1\\X_2\\ \vdots\\X_6\end{array}\right)
& \rightarrow &
\left(\begin{array}{c}X_2\\X_3\\ \vdots\\X_1\end{array}\right)
&\rightarrow &\cdots&\rightarrow& 
\left(\begin{array}{c}X_6\\X_1\\ \vdots\\X_5\end{array}\right)
&\rightarrow&
\left(\begin{array}{c}X_1\\X_2\\ \vdots\\X_6\end{array}\right)
\end{array}.  \label{orbitscorrespondence}
\end{eqnarray}
In the MSCA, all of the six eigenvalues of $M_6$ in the first set degenerate into a single value
$\Lambda \equiv [-2a(1-\varepsilon)]^6 \prod_{I=1}^6 X_I$  with degeneracy $\sum_{I=1}^6(N_I-1)=N-6$.
By (\ref{orbitscorrespondence}), (\ref{lineartransformation}) and (\ref{rconstraint})
we find $\Lambda=(-2b)^6 \prod_{i=1}^6 y_i$, that is, $\Lambda$ is nothing but the Lyapunov 
eigenvalue of the single logistic map for the $p=6$ motion.  
As for the other set, the $M_{\text{6; red}}$ for the symmetric configuration $\theta_I=1/6$ is 
a chain product of six matrices, that is,  $M^{6}_{\text{1; red}} M^{5}_{\text{1; red}}\cdots M^{1}_{\text{1; red}}$
with  
\begin{eqnarray}
M^{1}_{\text{1; red}}= -2a 
\left(\begin{array}{cccc}
(1- \varepsilon + \eta) X_1 &  \eta  X_2       &  \cdots  & \eta X_6 \\
\eta  X_1     & (1- \varepsilon + \eta) X_2   &  \cdots  & \eta X_6 \\
\vdots         & \cdots                         &   \ddots  & \vdots \\
\eta  X_1     & \cdots                         &  \cdots   &(1- \varepsilon + \eta) X_6
\end{array}
\right) , ~\eta=\frac{\varepsilon}{6} \label{msca6matrix}
\end{eqnarray}
and other five matrices are obtained by cyclically changing the orbit points $X_I$ by 
(\ref{orbitscorrespondence}).  By a simple algebra using (\ref{lineartransformation})
and (\ref{orbitscorrespondence}), we find that the eigenvalues of $M_6$ in the second 
set, which are in turn the ones of $M_{\text{6; red}}$, are $\Lambda$ with corrections of order $\eta$.\\
(4)
Now we are ready to work out the seagul cusp position. 
Because the $\Lambda$ is proportional to the product of the period six orbit points of the single logistic 
map $f_b$, it becomes zero when one of the orbit points becomes zero. 
At this very instance, the $N-6$ Lyapunov exponents 
become $-\infty$ and the other $6$ exponents become also very small proportionally to 
$\log(|\eta|)/6$. 
The $b$ is a solution of $f_b^6(0)=0$ and the relevant solution $b_c=1.772892$ gives
$\varepsilon_c=0.035192$ and $\lambda_{\text{max}}=-0.361519$ for $a=1.90$---
both are in remarkable agreement with the observed cusp of the GCML Lyapunov exponent.

Over the seagul $\varepsilon-$range, $M_{6;\text{ red}}$ has 
four complex [$(\lambda_k, \lambda_k^*),k=1,2)$] and two real eigenvalues
and gives four exponents. 
The $\lymax$ is given by one of the two sets of complex eigenvalues, while   
the $(N-6)$-fold degenerate exponent from $\Lambda$ runs in the mid of the four. 
The predicted $\lymax$ is shown in Fig.~\ref{lyapunovedep} and explains the data well.   
The slight deviation off the cusp is due the small population unbalance; it is the larger for the larger 
MSD events.  

{\it The dependence of the $\lambda_{max}$ on the population ratios}~~We proceed 
with the following algorithm after detecting the clusters by the gaps. 
The six MSCA${}^*$ 
clusters evolve in the bifurcated orbits of $p3c3$ MSCA. 
They can be regarded as three doublets --- $(C_{I_1},~C_{I_2}), ~I=1,2,3$ so that the two clusters 
$C_{I_1}$ and $C_{I_2}$ in a doublet evolve close together.
We combine the two populations in a doublet into one 
and define $s$, $t$ and $u$ as $(N_{I_1}+N_{I_2})/N$ in the decreasing order. 

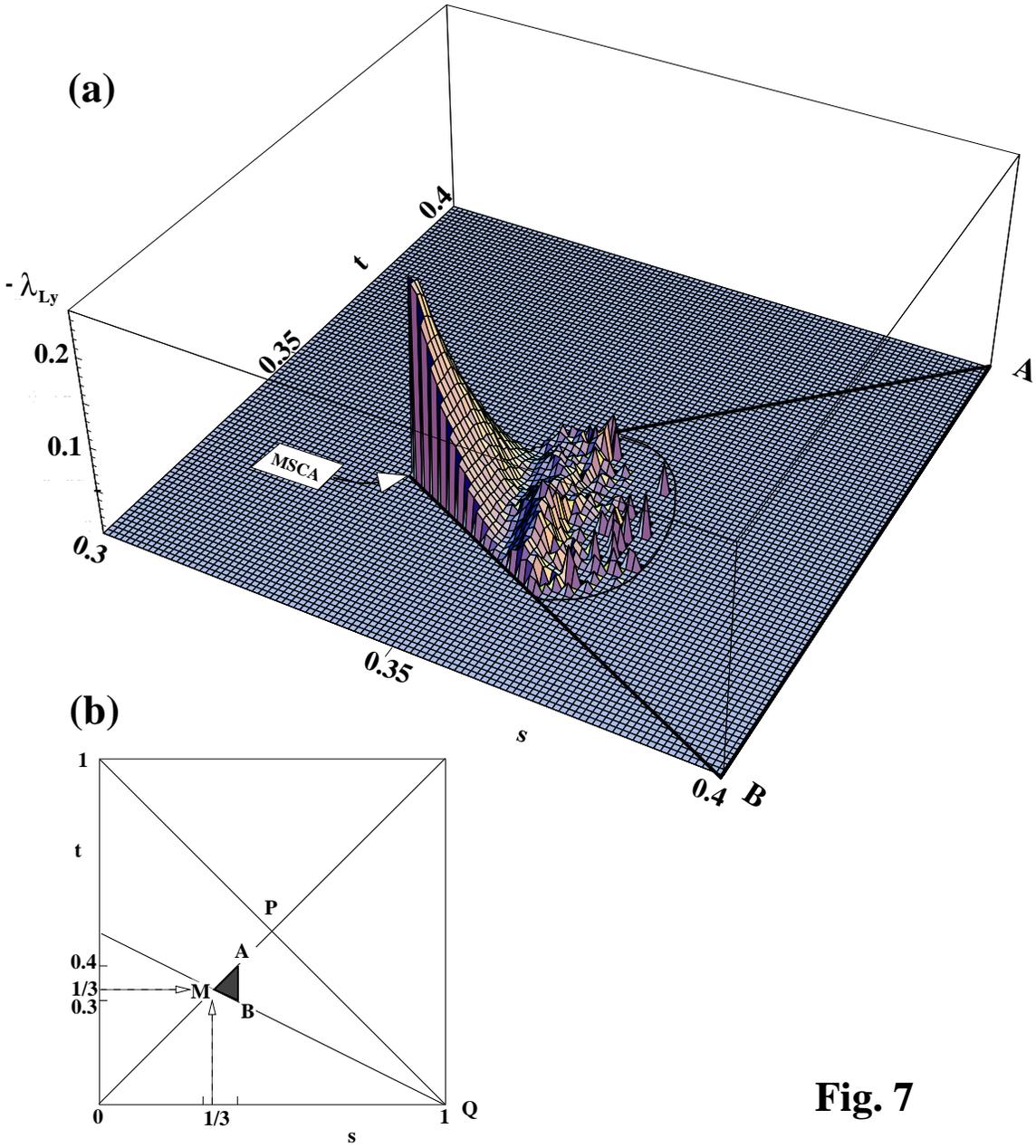
\begin{figure}
\caption{
(a) The maximum Lyapunov exponent of the bifurcated MSCA over the $s-t$ plane.
$\lambda_{\text{max}}$ for events in one $st-$bin are averaged,  
the sign is reversed (-$\lambda_{\text{max}}$) and the part $\lambda_{\text{max}}>0$ is truncated 
for a bird's eye view. $a=1.90$, $\varepsilon=0.035$, $N=10^4$ and totally $2 \times10^4$ events.
(b) The triangle $\Delta$PMQ is the allowed $(s,t)$ region by the constraint
$ s \ge t, ~~t \ge (1-s)/2, ~~1 \ge s+t $. The bifurcated MSCA events
accumulate in the top of the shaded tiny $\Delta$AMB, which is again a tiny part of
$\Delta$PMQ.
\label{p3c3pyramid}
}
\end{figure}

In Fig.~\ref{p3c3pyramid}(a), we exhibit the averaged $ - \lymax$ on the $s, ~t-$plane
from the $2 \times 10^4$ random events for $N=10^4$ GCML with $a=1.90$, $\varepsilon=0.035$.  
At the top of the pyramid-shaped surface the $\lymax$ is negative and at its minimum.
It occurs precisely at the most symmetric population configuration and we find
only an event with almost perfect population symmetry is formed.  
The $\lymax$ is negative over the bulk of events around the symmetric point --- MSCA is 
linearly stable. The exception occurs only near the boundary (the round curve),
where the $\lymax$ is mostly positive and small ($\lymax \lesssim 0.05$) and the maps form
the confined chaos.

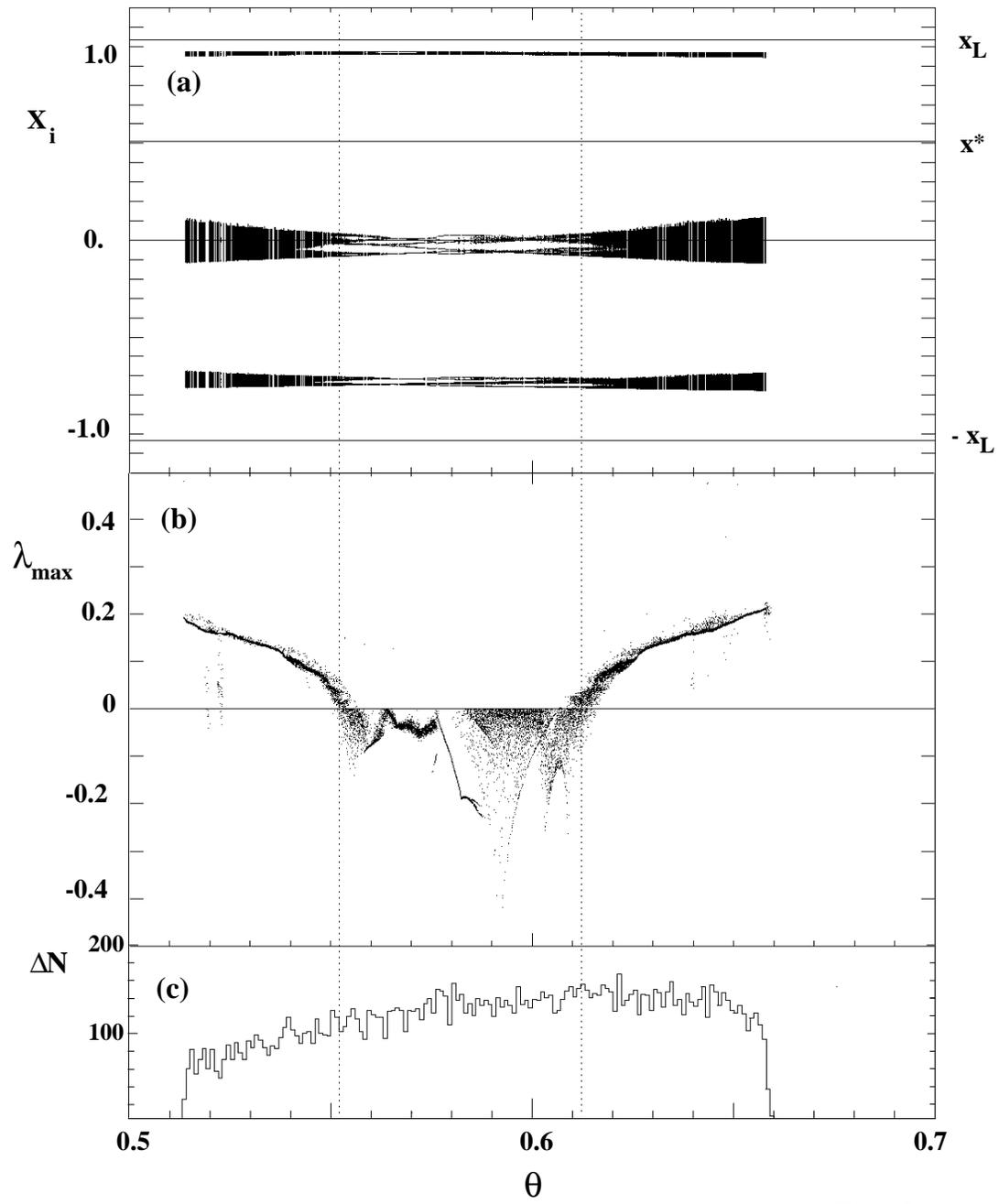
\begin{figure}
\caption{(a) The orbits of p3c2 cluster attractor for the last $512$ steps, 
(b) the maximum Lyapunov exponents, 
(c) the number of events in each bin ($\Delta\theta=10^{-3}$)
plotted at the population ratio $\theta$ for $18341$ events ($83\%$) 
with the stable cluster formation among totally $22000$ events.
$N=10^4$, $a=1.90$, $e=0.048$. 
Dotted lines to draw the eye and separate the tightly bound cluster states (central)
and confined chaos (two edges).
\label{p3c2window}
}
\end{figure}

%==================================================================
{\it The $p3c2$ Cluster Attractor}~~We have done a similar high statistics 
analysis at $a=1.90$, $\varepsilon=0.048$ for the same $N=10^4$ GCML. 
The final states are two fold; $p3c2$ cluster attractor ($83\%$) and 
the unstable $p3$ clusters with mixing (the rest). Hereafter we analyze the former 
in Fig.~\ref{p3c2window}.
In the region $0.55 \le \theta \le 0.61$, the $p3c2$ clusters are tightly bounded 
and linearly stable. Here the dynamics of the GCML is reduced to that of two clusters. 
Just like the $p2c2$ state in the ordered two clustered phase \cite{ka}, 
the $p3c2$ orbits bifurcate with the change of $\theta$ --- the ratio $\theta$ 
can be used as a control parameter even in the turbulent regime. 

However, there is a remarkable difference too.
In $p2c2$ there is no stable attractor for $\theta$ outside the window. 
In the turbulent regime, on the other hand, a loosely bound $p3c2$ state can be formed
 --- the three orbit bands in the edge regions.
This state is again the confined chaos. 
The $\lymax$ is positive ($0 \le \lymax \le 0.2$) and the maps fluctuate randomly 
in each of the two clusters. But the clusters are in a macroscopic period three motion. 
As the probability distribution shows, this is formed as frequently at the $p3c2$ 
cluster attractor. The state of confined chaos at the unbalanced population is 
a characteristic feature of the cluster attractors in the turbulent regime. 

%%%%%%%%%%%%%%%%
%%% Conclusion %%%%%%%
%%%%%%%%%%%%%%%%
\section{CONCLUSION}

In this article we have revisited the GCML of the logistic maps and studied in detail 
its so-called turbulent regime.  We have presented our new phenomenological findings 
in an extensive statistical analysis, which as a whole tell that the turbulent 
regime is under the systematic control of the periodic windows of the element logistic map. 
In particular we have shown that the hidden coherence occurs only in a very limited regions 
in the turbulent regime. 

There appears remarkable $p3c3$ MSCA states 
as well as $p3c2$ cluster attractors induced by the period three window of the element map. 
Our tuning condition predicts by a family of curves how the dynamics of the element 
map foliates in the parameter space of the GCML.  
It successfully explains the salient peak-valley structures of the MSD surface 
and tells us where to see the remarkable sequence of the cluster attractors of the type 
$p, c=p\rightarrow (p-1)\rightarrow (p-2) \rightarrow \cdots$.  

We have also investigated the linear stability of the period three cluster attractors. 
Both the $p3c3$ MSCA and its bifurcated state are linearly stable when the population 
symmetry is good and MSD of the meanfield is minimized.  
We have analytically explained the value of the coupling $\varepsilon$ at a given $a$ 
for the formation of the most stable bifurcated MSCA.  
The $p3c2$ cluster attractor is also linearly stable in the $\theta-$window even 
though the MSD of the $h(n)$ is quite high. 
For the unbalanced population configuration the system forms an interesting state 
of confined chaos which is a characteristic feature of the cluster attractors in the turbulent 
regime.

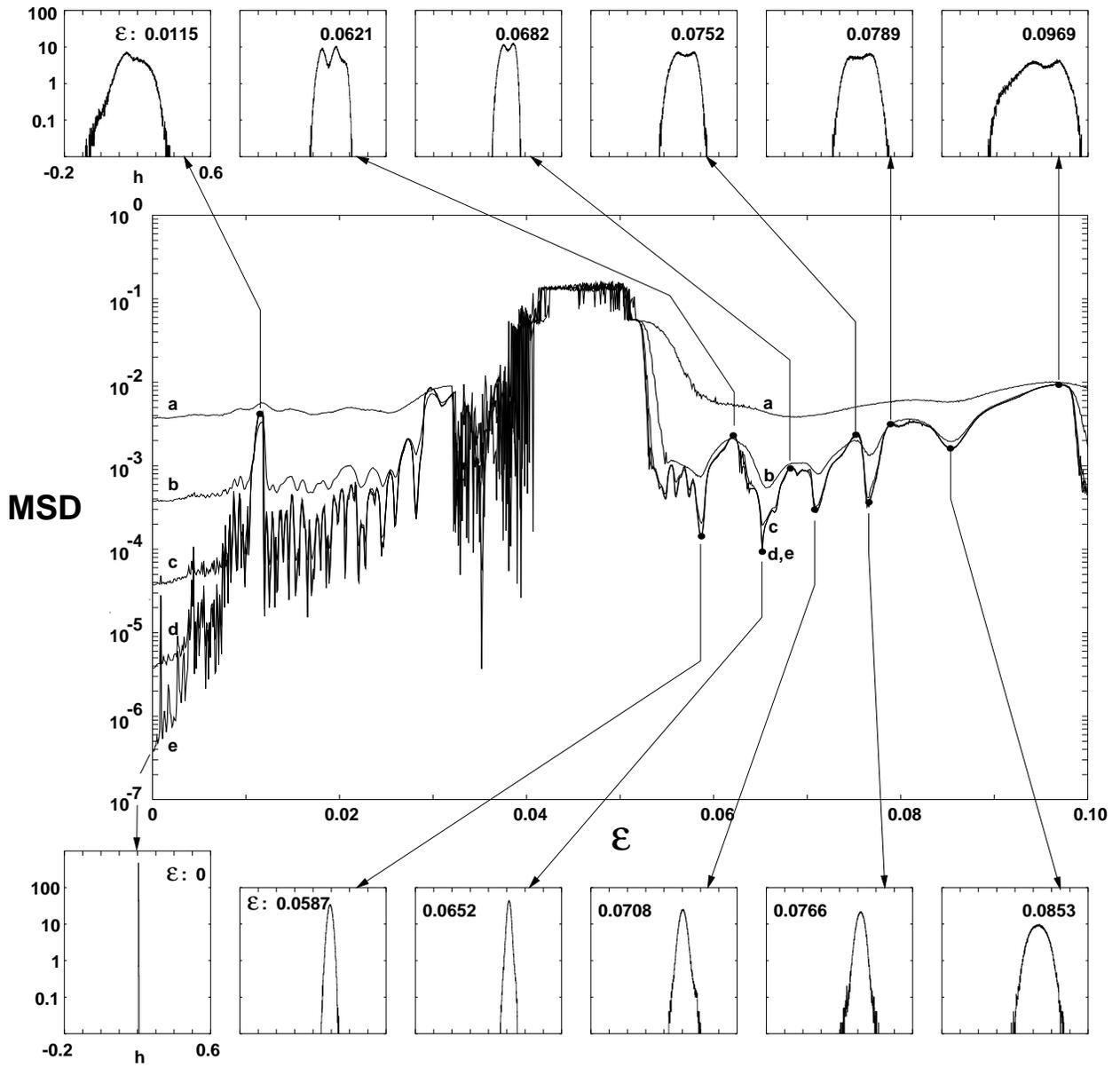
\begin{figure}
\caption{
Central diagram: MSD at $a=1.90$ as a function of $\varepsilon$
over $\varepsilon=0-0.10$ with inclement $10^{-4}$.
(a) $N=10^2$, (b) $10^3$,  (c) $10^4$,  (d)$10^5$, (e) $10^6$. 
Top boxes: The $h(n)$ distributions at MSD peaks. All are rank three. 
Bottom: the same at valleys. All are rank one and corresponds to the hidden coherence.  
\label{thermolimit}}\end{figure}

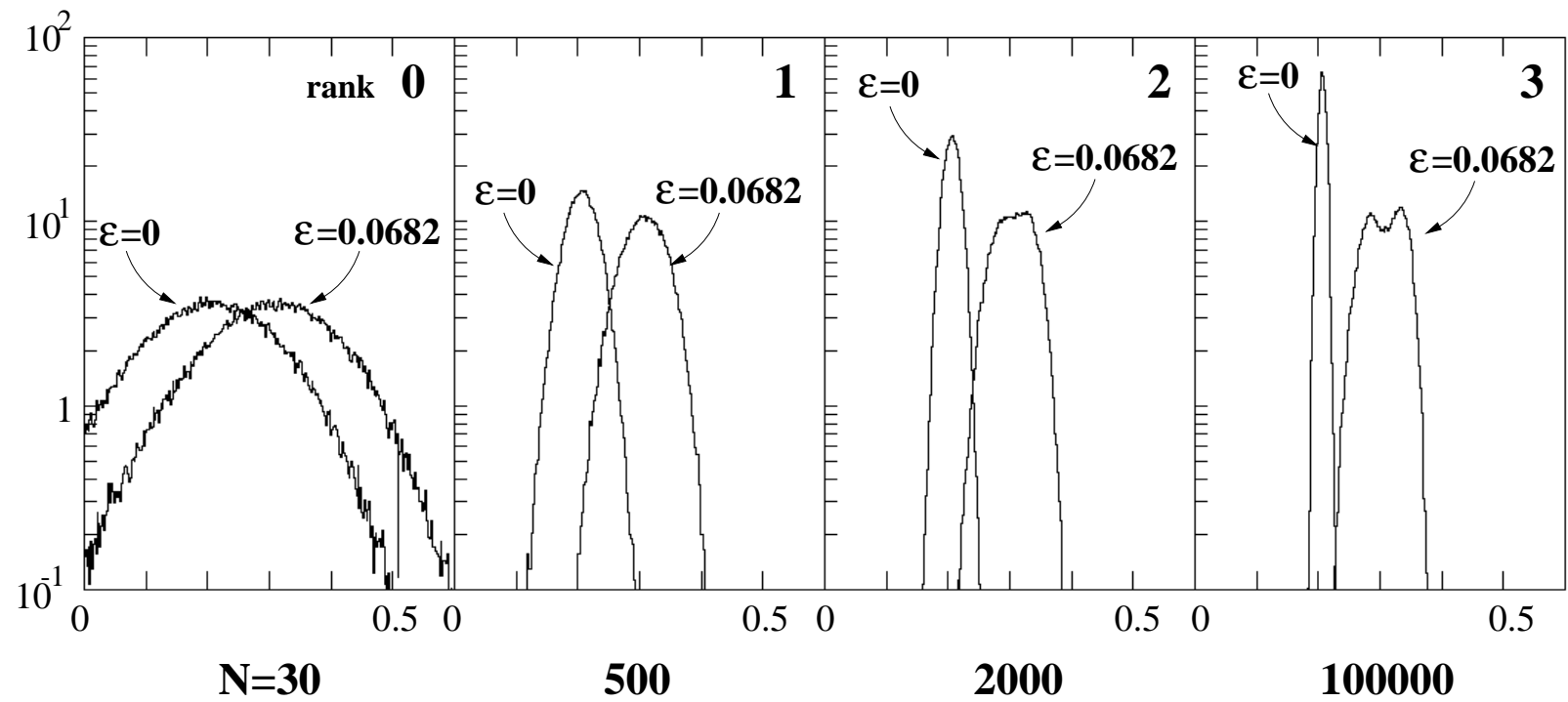
\begin{figure}
\caption{
The variation of the $h(n)$ distribution with the system size $N$ at 
$a=1.90, \varepsilon=0.0682$ and $N=3\times 10^1,5 \times 10^2,2 \times 10^3, 10^5$ from left to right.
Each is sampled in $10^5$ iterations from random start discarding 
the first $10^4$ steps and compared with a reference distribution (rank 0). 
The rank is given on top. 
\label{Ndepfig}
}
\end{figure}

There remain interesting unsolved problems.  
One concerns with the state of confined chaos newly found in the turbulent regime. 
It is a state consisting of a few clusters in macroscopic periodic motion
and maps move around chaotically inside each clusters. Regarding the linear stability 
the Lyapunov exponent is positive. It is tempting to single out the nonlinear effect which
confines the maps in periodic clusters. 
A related problem is the onset of the incomplete synchronization with the decrease of 
the reduction factor $r$ along the foliation curves. 
The other concerns with the variation of the dynamics with the system size $N$. 
We have found that the system becomes an extremely sensitive mirror of the element dynamics 
with increasing $N$. The salient evidences are shown in Fig.~\ref{thermolimit} and Fig.~\ref{Ndepfig}, 
but we are unable to explain why so. 
In field theory the vacuum at the spontaneous break down of the symmetry
is stable only when the degree of dynamical degree of freedom is infinite \cite{weinberg}.
If we may regard the randomness of GCML maps as a symmetry, the MSCA with no $h(n)$ 
fluctuation corresponds to a vacuum at the symmetry breakdown and the formation of it by 
synchronization the onset of the ordered parameter. 
The resolution of the finite size effect in GCML is so tempting since 
it may bridge the synchronization of the maps and onset of the order parameter  in the field theory 
in quantitative terms. 

As a whole this work is an exploration of order in the chaos and we have found that the turbulent 
regime of GCML is controlled by the foliation of the single logistic dynamics.

%%%%%%%%%%%%%%%%
%%% Acknowledgement %%%
%%%%%%%%%%%%%%%%
 
\acknowledgments
It is our pleasure to thank Hayato Fujigaki,  Fumio Masuda,  Ko-ichi Nakamura, Maki Tachikawa, 
Norisuke Sakai,  Wolfgang Ochs and Hidehiko Shimada  for useful discussions and encouragement.

Some observations were partially reported in earlier articles by one of us (TS) \cite{early} 
which include the finding of the manifestation of $p3c3$ MSCA and $p3c2$ attractor state.
While preparing the final manuscript to include results on turbulent regime at huge $N$ and 
on the stability of MSCA, we have noticed related works on the foliation of the logistic windows.  
A. P. Parravano and M. G. Cosenza have independently reported MSCA\cite{parravano}. 
T. Shibata and K. Kaneko have also independently found the foliation of windows in the mean 
field fluctuations and called it as a tongue structure \cite{shibata}.  
Both parallel works\cite{parravano,shibata} overlap ours with respect to 
the foliation but neither the manifestation of the attractors of the type $p>c$ nor the stability 
of MSCA were discussed in them.
 
This work  was supported  by the Faculty Collaborative Research Grant from Meiji University, 
Grant-in-Aids for Scientific Research from Ministry of Education, 
Science and Culture of Japan, and Grant for High Techniques Research from both organizations.

%%%%%%%%%%%%%%%%
%%% References %%%%%%
%%%%%%%%%%%%%%%%

%%-----------------------------------------------------------------------------------------------------%%

%%%%%%%%%%%%%%%%
%%% Figure Captions %%%%
%%%%%%%%%%%%%%%%

%---Fig. 1 ---MSD surface and rank plot

%--- Fig. 2 ---thermolimit

%--- Fig. 3 ---Ndepfig(4 MSD at N=30-10^5)

%--- Fig. 4---Traversing the Turbulent Regime:

%--- Fig. 5--- Foliation of the single map dynamics

%--- Fig. 6 ---Lyapunov and MSD e-dependence

%--- Fig.  7 ---p3c3pyramid

%--- Fig.  8 ---p3c2 window

%%%%%%%%%%%%%%%%
%%%  Tables     %%%%%%%
%%%%%%%%%%%%%%%%
\mediumtext
%----------------------------------------------------------------------------------------------
\begin{table}
\caption{
The $p=3$ periodicity manifestation in the turbulent regime. $a=1.90$, $N=10^4$-$10^6$. 
\label{tablep3c3}}
\begin{tabular}{cccc}
$\varepsilon-$range&Prediction&MSDsurface&state\\
\tableline
$0.032-0.0352\tablenotemark[1]-0.037$&0.0305-0.035192-0.0363&deep valley&
bifurcated $p3c3$~MSCA\\
$0.037-0.041$&0.0363-0.0422&lower band&$p3c3$~MSCA\\
$0.037-0.041$&&upper band&$p3c2$ cluster attractor\\
$0.041-0.050$&0.0422-0.0514&prominent peak&$p3c2$ cluster attractor\\
\end{tabular}
\tablenotemark[1]{The downwards cusp position of the MSD valley.  }
\end{table}
%-----------------------------------------------------------------------------------------------
\begin{table}
\caption{
The outstanding windows and their $b$, $y^*(b)$ values. 
\label{tableabcd}}
\begin{tabular}{cccccc}
    period & A: intermittency  \tablenotemark[1] 
 & B: lower threshold& C: the first bifurcation  &D: closing point & Width\tablenotemark[2]  \\
\tableline
7     &1.5740, 0.3943&1.5748, 0.3857  &1.5754, 0.3846  &1.5762, 0.3847 & 0.0014  \\   
5     &1.6220, 0.3610&1.6244, 0.3077  &1.6284, 0.3012  &1.6333, 0.3032  &0.0089\\   
7     &1.6735, 0.3189&1.6740, 0.2676  &1.6744, 0.2678  &1.6749, 0.2677 & 0.0009\\   
\tableline
3     &1.7350, 0.2836&1.7500, 0.0952&1.7685, 0.0685  &1.7903, 0.0800 &0.0403  \\   
\tableline
5     &1.8597, 0.1823&1.8606, 0.0984  &1.8614, 0.0990  &1.8623, 0.0987 &0.0017  \\   
4     &1.9390, 0.1287&1.9406, -0.1633  &1.9415, -0.1668  &1.9427, -0.1657 &0.0021  
\end{tabular}
\tablenotemark[1]{The starting point of intermittency. 
 }\\
\tablenotemark[2]{$\Delta b\equiv b_B - b_D$ 
 }
\end{table}
%-----------------------------------------------------------------------------------------------------
\begin{table}
\caption{Samples of the sequences of attractors with $p \ge c$.  $N=10^4$ and $a$ in parentheses.
\label{tableotherpc}}
\begin{tabular}{cllr}
$p=4$: ~$b:1.9406-1.9427\tablenotemark[1]$&( $a=1.95, ~r\approx 0.995)$\\
\tableline
%---------------------------------------------------------------------------------------------------
$c$&~~4~(MSCA)&~~~~~~~~3&2~~~~~~~~~~\\
%------------------------------------------------------------------------------------------------
$\varepsilon$&$0.0019-0.0022$&$0.0022 -0.0024$&$0.0024-0.0026$\\ 
$\varepsilon_{pred}$\tablenotemark[2]&$0.0020-0.0022$&$0.0022----$&$---0.0030$\\ 
\tableline
$p=5$: ~$b:1.6244-1.6333\tablenotemark[1]$&( $a=1.66,~ r \approx 0.980$ )\\
\tableline
%------------------------------------------------------------------------------------------------------------
$c$  &~~5~(MSCA)\tablenotemark[3]&~~~~~~4&3~~~~~~~\\
%------------------------------------------------------------------------------------------------------------
$\varepsilon$&$0.00986-0.0118$&$0.0114-0.0124$&$0.0124-0.0130$\\
$\varepsilon_{pred}$\tablenotemark[4]&$0.00950-0.0112$&$0.0112---$&$---0.0140$\\
\end{tabular}
\tablenotemark[1]{The range of the window from B to D, i.e. intermittent region not included.}\\
\tablenotemark[2]{The $\varepsilon-$range predicted by the tuning condition and 
the window data in Table \ref{tableabcd}. The predictions are from C-B for MSCA and B-A for $p>c$ states.} \\
\tablenotemark[3]{For most events, the attractor is either bifurcated or consists of five bands. }\\
\tablenotemark[4]{D-C for MSCA and C-A for $p>c$ cluster attractor taking into account the bifurcation.} 
\end{table}

\begin{references}

\bibitem{ka}
K. Kaneko, Phys. Rev. Lett. {\bf 63}, 219 (1989).

\bibitem{kb1}
%Law of Large Numbers 
K. Kaneko, Phys. Rev. Lett. {\bf 65}, 1391 (1990).
% extended and a little on FP

\bibitem{pikovskyb}
A. S. Pikovsky, and J. Kurths, Phys. Rev. Lett. {\bf 72}, 1644 (1994).

\bibitem{kb2}
K. Kaneko, Physica (Amsterdam) {\bf D55}, 368 (1992); ibid., {\bf 86}, 158 (1995).
% detailed FP, and tent

\bibitem{kc}
K. Kaneko, Physica (Amsterdam) {\bf D34}, 1 (1989);  
J. P. Crutchfield and K. Kaneko, in {\it Directions in Chaos}, 
edited by B. -L. Hao (World Scientific, Singapore, 1987).
%cml

\bibitem{kd}
K. Kaneko, Physica (Amsterdam) {\bf 41D}, 137 (1990).
% Coding

\bibitem{sinha}
K. Kaneko,  Physica  {\bf 54D}, 5 (1991); 
S. Sinha, D. Biswas, M. Azam, and S. V. Lawande, Phys. Rev.  {\bf A46}, 3193 (1992). 
%circle map 

\bibitem{pecora}
L. M. Pecora and T. L. Carroll, Phys. Rev. Lett. {\bf 64}, 821 (1990); 
T. L. Carroll and L. M. Pecora, Physica {\bf 67}, 126 (1993).

\bibitem{kowalski}
J. M. Kowalski and G. L. Albert, Phys. Rev. {\bf A42}, 6260 (1990).

\bibitem{noise}
A. Maritan and J. R. Banavar, Phys. Rev. Lett. {\bf 72}, 1451 (1994).

\bibitem{fns}
H. Fujigaki, M. Nishi, and T. Shimada, Phys. Rev. {\bf E53}, 3192 (1996);    
H. Fujigaki and T. Shimada, Phys. Rev. {\bf E55}, 2426 (1997).  

\bibitem{pikovskya}
A. S. Pikovsky, M. G. Rosenblum, and J. Kurths, Europhys. Lett. {\bf 34}, 165 (1996); 
M. G. Rosenblum, A. S. Pikovsky, and J. Kurths, Phys. Rev. Lett. {\bf 76}, 1804 (1996).
%phase entrainment 

%\bibitem{yorke}
%E. Ott, C. Grebogi, and J. A. Yorke, Phys. Rev. Lett. {\bf 64}, 1196 (1990).

\bibitem{biology}
J. J. Hopfield, Proc. Natl. Acad. Sci. USA (Biophysics){\bf 79}, 2554 (1982);
W.A. Little, Math. Biosci. {\bf 19}, 101 (1974);
D. J. Amit, H. Gutfreund and H. Sompolinski, Phys. Rev. {\bf A32}, 1007 (1985).

\bibitem{spinglass}
M. M\'ezard, G. Parisi, and M. A. Virasoro (Editors), 
{\it Spin glass theory and beyond} (World Scientific, 1987).


\bibitem{Ruelle}
A. Lasota and M. C. Mackey, {\it Chaos, Fractals, and Noise} (Springier, 1994), Chap. 3. 2 and 7. 4. 
%D. Ruelle, {\it Thermodynamic Formalism} (Addison Wesley, Reading, MA,1978);
Y. Oono and Y. Takahashi,  Prog. Theor. Phys.{\bf  63}, 1804 (1980); 
%H. H.  Hasegawa and W. C. Saphir,  Phys. Rev. {\bf A46}, 7401 (1992);
W. Just, J. Stat. Phys. 79, 429 (1995).

\bibitem{ershov}
S. V. Ershov and A. B. Potapov, Physica {\bf D86}, 523 (1995).

\bibitem{PerezCerdeira}
G.Perez and H. A. Cerdeira, Phys. Rev. {\bf A46}, 7492 (1992).

%\bibitem{shimadakikuchi}
%T. Shimada and K. Kikuchi, in preparation.

\bibitem{nagashima}
I. Shimada and T. Nagashima, Prog. Theor. Phys. {\bf 61}, 1605 (1979); 
G. Benettin, L. Galgani, A. Giorgilli, and J. M. Strelcyn, C. R. Acad.Sci.Paris {\bf A14} , A-431 (1978).

%\bibitem{oseledec}
%V. I. Oseledec, Trans. Moscow Math. Soc. 19, 197 (1968).

%\bibitem{benettin}
%G. Benettin, L. Galgani, and J. M. Strelcyn, Phys. Rev. {\bf A14}, 2338 (1976).

\bibitem{weinberg} See, e.g. S. Weinberg, {\it The Quantum Theory of Fields}, 
(Cambridge, 1996) vol. 2.  
A macroscopic broken rotational symmetry state realized by a chair is discussed in p. 195. 

\bibitem{early} 
T. Shimada,  Technical Report of IEICE, NLP97-159,  71 (1998);
and an early version of this paper, chao-dyn/9810007.

\bibitem{parravano}
A. P. Parravano and M. G. Cosenza, Cluster dynamics in systems with constant mean 
field coupling, chao-dyn/9808009.

\bibitem{shibata}
T. Shibata and K. Kaneko, 
% On the Tongue-Like Bifurcation Structure of the Mean-Field Dynamics in a Network of Chaotic Elements, 
Physica {\bf D}124, 177 (1998).
\end{references}
\end{document}